\documentclass[preprint2]{proto}
\usepackage{times}

\def\micron{$\mu$m}

\newcommand\be{\begin{equation}}
\newcommand\en{\end{equation}}

\begin{document}

\title{\textbf{\LARGE An Observational Perspective of
Transitional Disks}}

\author {\textbf{\large Catherine Espaillat}} 
\affil{\small\em Boston University}

\author {\textbf{\large James Muzerolle}} 
\affil{\small\em Space Telescope Science Institute}

\author {\textbf{\large Joan Najita}} 
\affil{\small\em National Optical Astronomy Observatory}

\author {\textbf{\large Sean Andrews}} 
\affil{\small\em Harvard-Smithsonian Center for Astrophysics}

\author {\textbf{\large Zhaohuan Zhu}} 
\affil{\small\em Princeton University}

\author {\textbf{\large Nuria Calvet}} 
\affil{\small\em University of Michigan}

\author {\textbf{\large Stefan Kraus}} 
\affil{\small\em University of Exeter}

\author {\textbf{\large Jun Hashimoto}} 
\affil{\small\em The University of Oklahoma}

\author {\textbf{\large Adam Kraus}} 
\affil{\small\em University of Texas}

\author {\textbf{\large Paola D'Alessio}} 
\affil{\small\em Universidad Nacional Aut{\'o}noma de M{\'e}xico}

\begin{abstract}
\baselineskip = 11pt
\leftskip = 0.65in 
\rightskip = 0.65in
\parindent=1pc
{\small 
Transitional disks are objects whose inner disk regions have undergone substantial clearing.
The {\it Spitzer Space Telescope} produced detailed spectral energy distributions (SEDs) of transitional disks that allowed us to infer their radial dust disk structure in some detail, revealing the diversity of this class of disks.  The growing sample of transitional disks also opened up the possibility of demographic studies, which provided unique insights.
There now exist (sub)millimeter and infrared images that confirm the presence of large clearings of dust in 
transitional disks.  In addition, protoplanet candidates have been detected within some of these clearings. 
Transitional disks are thought to be a strong link to planet formation around young stars and are a key area to study if further progress is to be made on understanding the initial stages of planet formation. 
Here we provide a review and synthesis of transitional disk observations to date with the aim of
providing timely direction to the field, which is about to undergo its next burst of growth as ALMA reaches its full potential.  
We discuss what we have learned about transitional disks from SEDs,
color-color diagrams, and imaging in the (sub)mm and infrared.  We note the limitations of
these techniques, particularly with respect to the sizes of the clearings currently detectable, and
highlight the need for pairing broad-band SEDs with multi-wavelength images to paint a more detailed picture of transitional disk structure.
We review the gas in transitional disks, keeping in mind that future observations
with ALMA will give us unprecedented access to gas in disks, and also observed infrared variability 
pointing to variable transitional disk structure, which may have implications for disks in general.
We then distill the observations into constraints for the main disk clearing mechanisms proposed to date (i.e., photoevaporation, grain growth, and companions) and explore how the expected observational signatures
from these mechanisms, particularly planet-induced disk clearing, 
compare to actual observations.
Lastly, we discuss
future avenues of inquiry to be pursued with ALMA, JWST, and next generation of ground-based telescopes.
\\~\\~\\~\\~}

\end{abstract}  

\section{\textbf{INTRODUCTION}}
\bigskip
\bigskip
Disks around young stars are thought to be the sites of planet formation. However, many questions exist concerning how the gas and dust in the disk evolve into a planetary system. 
Observations of T Tauri stars (TTS) may provide insights into these questions, and a subset of TTS, the ``transitional disks,'' have gained increasing attention in this regard. The unusual SEDs of transitional disks (which feature infrared excess deficits) may indicate that they have developed significant radial structure. 

Transitional disk SEDs were first identified by \citet{strom89,skrutskie90} from near-infrared (NIR) ground-based photometry and IRAS mid-infrared (MIR) photometry. These systems exhibited small NIR and$/$or MIR excesses, but significant mid- and far-IR (FIR) excesses indicating that the {\it dust} distribution of these
disks had an inner hole (i.e., a region that is mostly devoid of small dust grains from a radius R$_{hole}$ down to the central star). They proposed that these disks were in transition from objects with optically thick disks that extend inward to the stellar surface (i.e., Class II objects) to objects where the disk has dissipated (i.e., Class III objects), possibly as a result of some phase of planet formation.
A few years later, \citet{marsh92, marsh93}
proposed that such transitional disk SEDs were consistent with the expectations
for a disk subject to tidal effects exerted by companions, either stars or planets.

More detailed studies of transitional disks became possible as
increasingly sophisticated instruments became available. 
The spectrographs on board {\em ISO} were able to
study the brightest stars and \citet{bouwman03}
inferred a large hole in the disk of the Herbig Ae$/$Be star HD~100546 based on its SED.
Usage of the term ``transitional disk'' gained substantial momentum in the literature
after the {\it Spitzer Space Telescope}'s \citep{werner04} 
Infrared Spectrograph \citep[IRS;][]{houck04}
was used to study disks with inner holes \citep[e.g.,][]{dalessio05,calvet05}.  
{\it Spitzer} also detected disks with an annular ``gap'' within the disk as opposed to holes \citep[e.g.,][]{brown07,espaillat07b}. 
In this review, we use the term {\it transitional disk} to
refer to an object with an inner disk hole and {\it pre\nobreakdashes-transitional disk} to refer to
a disk with a gap.
For many (pre\nobreakdashes-)transitional disks, the inward truncation of the
outer dust disk has been confirmed, predominantly through (sub)millimeter
interferometric imaging \citep[e.g.,][]{hughes07,brown08,hughes09,brown09,andrews09,isella10b,andrews11}. 
We note that (sub)mm imaging is not currently capable of distinguishing between a hole or gap in the disk 
(i.e., it can only detect a generic region of clearing or a ``cavity'' in the disk).
Also, it has not yet been confirmed if these clearings detected in the dust disk are present in the gas disk as well.  The combination of these dust cavities
with the presence of continuing gas accretion onto the central star is a challenge to theories
of disk clearing.

The distinct SEDs of (pre\nobreakdashes-)transitional disks have lead many researchers to
conclude that these objects
are being caught in an important phase in disk evolution.  One
possibility is that these disks are forming planets given that
cleared disk regions are predicted by theoretical planet formation models
\citep[e.g.,][]{paardekooper04, zhu11,dodson11}.  Potentially supporting this, there exist observational reports
of protoplanet candidates in (pre\nobreakdashes-)transitional disks \citep[e.g., LkCa~15, T~Cha;][]{kraus11,huelamo11}.
Stellar companions can also clear the inner disk \citep{artymowicz94} but many stars harboring
(pre\nobreakdashes-)transitional disks are single stars \citep{kraus11}.  Even if companions are
not responsible for the clearings seen in all (pre\nobreakdashes-)transitional disks, these objects still have
the potential to inform our understanding of how disks dissipate,
primarily by providing constraints for disk clearing models involving photoevaporation
and grain growth.

In this chapter, we will review the key observational constraints on the dust and gas properties of (pre\nobreakdashes-)transitional disks and examine these in the context of theoretical disk clearing mechanisms. 
In \S 2, we look at SEDs (\S~2.1) as well as (sub)mm (\S~2.2) and IR (\S~2.3) imaging.
We also review IR variability in (pre\nobreakdashes-)transitional disks (\S~2.4) and gas observations (\S~2.5).
In \S 3, we turn the observations from \S 2 into constraints for the main disk clearing mechanisms proposed to date (i.e., photoevaporation, grain growth, and companions) and discuss these mechanisms in light of these constraints.
In \S 4, we examine the demographics of (pre\nobreakdashes-)transitional disks (i.e., frequencies, timescales, disk masses, accretion rates, stellar properties) in the context of disk clearing and in \S 5 we conclude with possibilities for future work in this field.
\bigskip
\bigskip
\bigskip
\section{\textbf{OVERVIEW OF OBSERVATIONS}}
\bigskip

\bigskip
In the two decades following \citet{strom89}'s identification of the first transitional disks using NIR and MIR photometry, modeling of the SEDs of these disks, enabled largely by {\it Spitzer} IRS, inferred the presence of holes and gaps that span several AU (\S~2.1).  Many of these cavities were confirmed by (sub)mm interferometric imaging (\S~2.2) and IR polarimetric and interferometric (\S~2.3) images.  Later on, MIR variability was detected which pointed to structural changes in these disks (\S~2.4).  While there are not currently as many constraints on the gas in the disk as there are for the dust, it is apparent that the nature of gas in the inner regions of (pre\nobreakdashes-)transitional disks differs from other disks (\S~2.5).  These observational results have significant implications for our understanding of planet formation and we review them in the following sections.

\begin{figure}[ht!]
\begin{center}
\includegraphics[width=5.4cm]{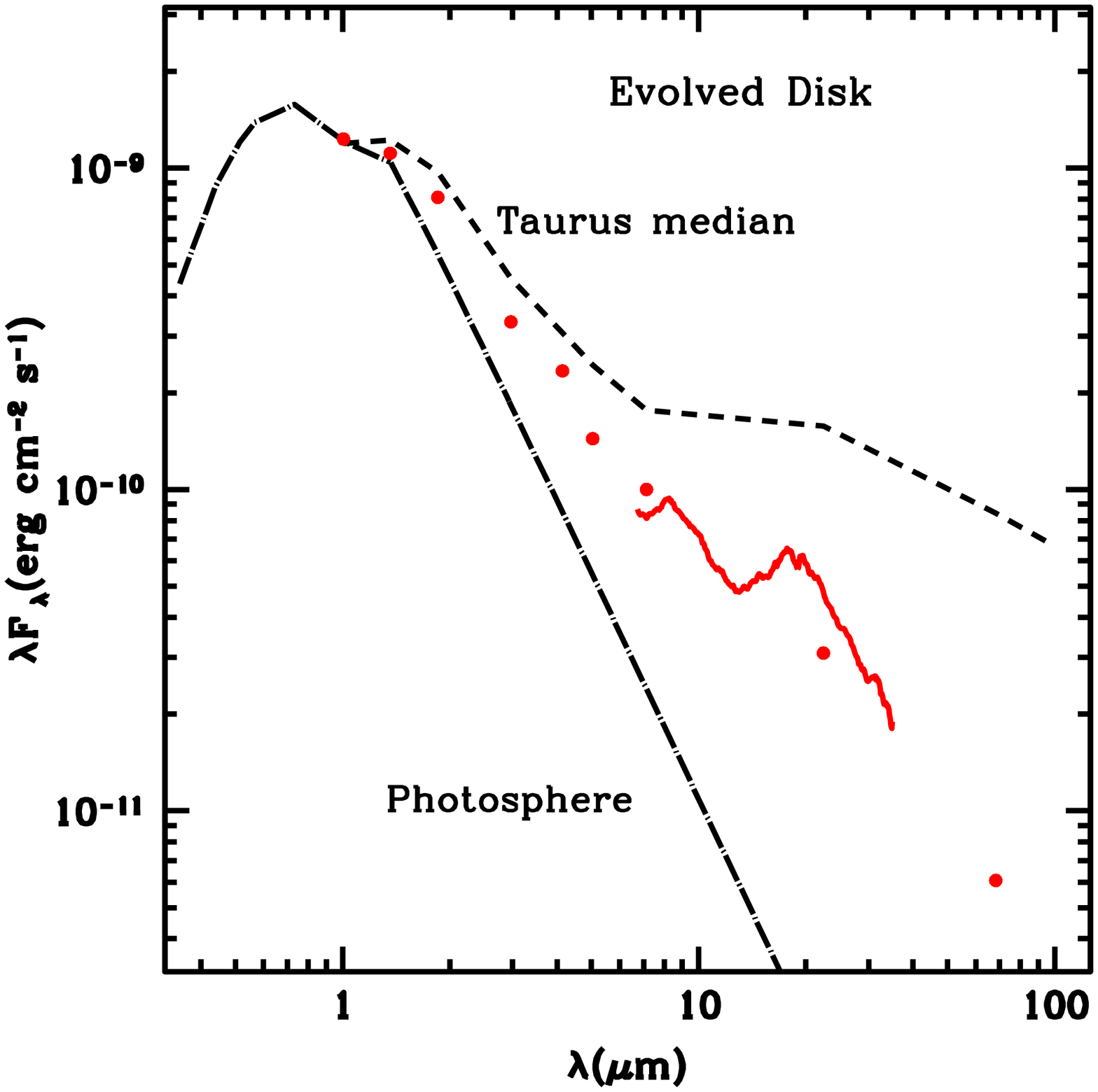}\\
\includegraphics[width=5.4cm]{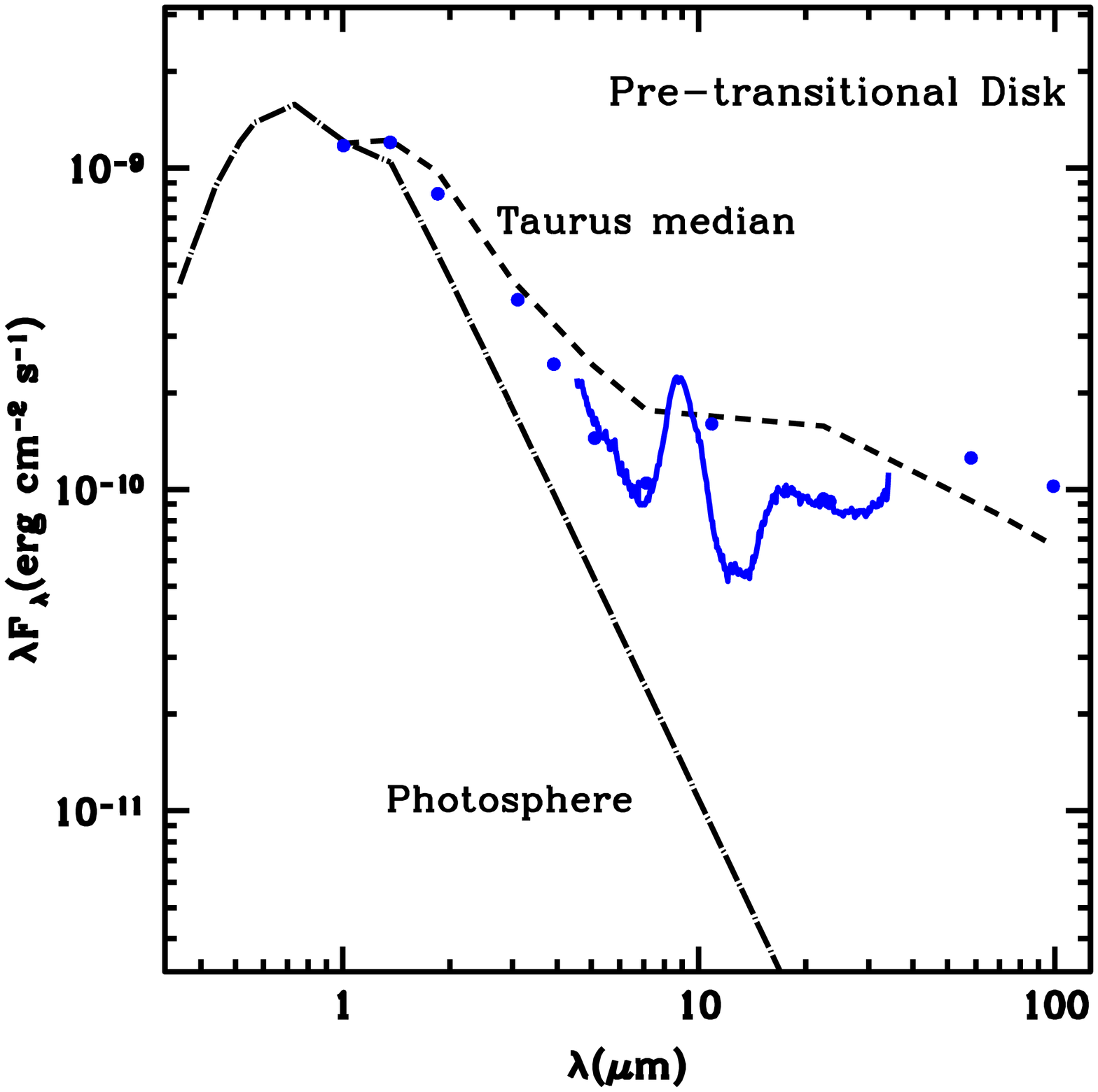}\\
\includegraphics[width=5.4cm]{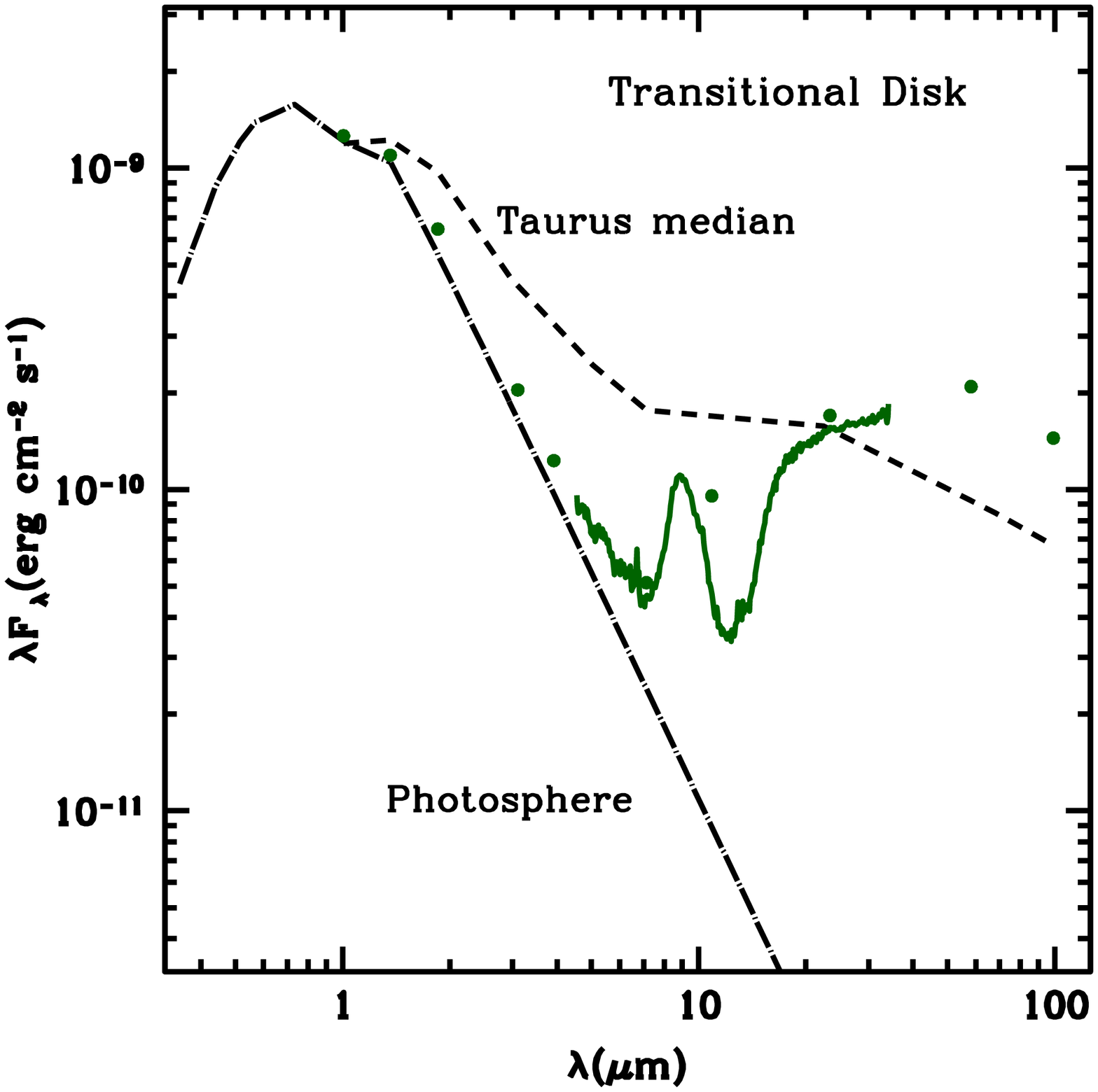}
\end{center}
 \vspace{-0.3 cm}
 \caption{\small 
SEDs of 
an evolved disk \citep[top; RECX 11;][]{ingleby11b},
a pre\nobreakdashes-transitional disk \citep[middle; LkCa 15;][]{espaillat07b},
and a transitional disk \citep[bottom; GM Aur;][]{calvet05}.
The stars are all K3-K5 and
the fluxes have been corrected for reddenning and
scaled to the stellar photosphere (dot-long-dashed line) for comparison. 
Relative to the Taurus median \citep[short-dashed line;][]{dalessio99},
an evolved disk has less emission at all wavelengths,
a pre\nobreakdashes-transitional
disk has a MIR deficit (5--20~{\micron}, ignoring the 10~{\micron} silicate emission feature), but comparable
emission in the NIR (1--5~{\micron}) and at longer wavelengths, and
a transitional disk has a deficit of emission in the NIR and MIR with comparable emission
at longer wavelengths.
}  
\label{figseds}
\end{figure}

\begin{figure}[ht!]
\begin{center}
\includegraphics[width=6.2cm]{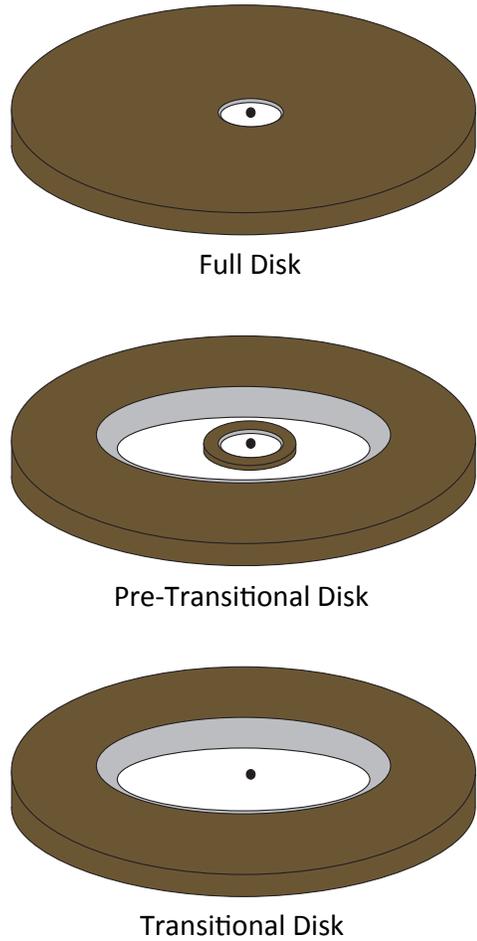}\\
 \vspace{-0.3 cm}
\end{center}
\caption{\small 
Schematic of full (top), pre\nobreakdashes-transitional (middle), and transitional (bottom) disk structure. For the full disk, progressing outward from the star (black) is the inner disk wall (light gray) and outer disk (dark brown).  Pre\nobreakdashes-transitional disks have an inner disk wall (light gray) and inner disk (dark brown) followed by a disk gap (white), then the outer disk wall (light gray) and outer disk (dark brown).  The transitional disk has an inner disk hole (white) followed by an outer disk wall (light gray) and outer disk (dark brown). 
}  
\label{figsch2}
\end{figure}

\bigskip
\noindent
\textbf{ 2.1 Spectral Energy Distributions}
\bigskip

SEDs are a powerful tool in disk studies as they provide information over a wide range of wavelengths, tracing different emission mechanisms and material at different stellocentric radii.
In a SED, one can see the signatures of gas accretion 
\citep[in the ultraviolet; see PPIV review by][]{calvet00}, the stellar photosphere 
(typically $\sim$1~{\micron} in TTS), and the dust in the disk (in the IR and longer wavelengths).  
However, SEDs are not spatially resolved and this information must be supplemented by
imaging, ideally at many wavelengths (see \S~2.2--2.3).
Here we review what has been learned from studying the SEDs
of (pre-)transitional disks, particularly using {\it Spitzer} IRS, IRAC, and MIPS.

\bigskip
\noindent
\textit{SED classification}
\medskip

A popular method of identifying transitional disks is to
compare individual SEDs to the median SED of disks in the 
Taurus star-forming region (Fig.~\ref{figseds}, dashed line in panels).  
The median Taurus SED is typically taken
as representative of an optically thick full disk (i.e., a disk with no 
significant radial discontinuities in its dust distribution).
The NIR emission (1--5~{\micron}) seen in the SEDs of full disks is dominated by the 
``wall'' or inner edge of the dust disk.  This wall is located 
where there is a sharp change at the radius at which the dust
destruction temperature is reached and dust sublimates.
TTS in Taurus have NIR excess emission which can be fit by
blackbodies with temperatures within the observed
range of dust sublimation temperatures \citep[1000--2000 K;][]{monnier02},
indicating that there is optically thick material located at the dust destruction
radius in full disks \citep{muzerolle03}.  
Roughly, the MIR emission in the SED traces the inner tens of AU in disks
and emission at longer wavelengths comes from outer radii of the
disk.  

The SEDs of transitional disks are characterized by
NIR (1--5~{\micron}) and MIR emission (5--20~{\micron}) similar to that of a stellar photosphere,
while having excesses at wavelengths $\sim$20~{\micron} and beyond comparable to the Taurus median,
\citep[Fig.~\ref{figseds}, bottom;][]{calvet05}. 
From this we can infer that the small, hot dust that typically emits at these wavelengths in full disks has been removed
and that there is a large hole in the inner disk, larger than can be explained by dust sublimation (Fig.~\ref{figsch2}, bottom).
Large clearings of dust in the submm regime have been identified in disks
characterized by this type of SED \citep[\S~2.2; e.g.,][]{hughes09,brown09,andrews11}, 
confirming the SED interpretation. 
We note that disks with holes have also been referred to as cold disks \citep{brown07} or weak excess transitional disks \citep{muzerolle10}, but
here we use the term ``transitional disks.''

A subset of disks with evidence of clearing in the submm
show significant NIR excesses relative to their stellar photospheres, in some cases comparable to the median Taurus SED, but still exhibit MIR
dips and substantial excesses beyond $\sim$20~{\micron} (Fig.~\ref{figseds}, middle).
This NIR excess is blackbody-like with temperatures
expected for the sublimation of silicates \citep{espaillat08a}, similar to the NIR excesses 
in full disks discussed earlier.
This similarity indicates that these disks
still have optically thick material close to the star, possibly a remnant
of the original inner disk, and these gapped disks have been dubbed pre\nobreakdashes-transitional
disks \citep{espaillat07b}, cold disks \citep{brown07}, or warm transitional disks \citep{muzerolle10}.  Here we adopt the term ``pre\nobreakdashes-transitional
disks'' for these objects.  In Table~1 we summarize some of properties of many of the well-known
(pre\nobreakdashes-)transitional disks.

We note that some SEDs have
emission that decreases steadily at all wavelengths (Fig.~\ref{figseds}, top).
These disks have been called a variety of names:
anemic \citep{lada06},
homologously depleted \citep{currie11},
evolved \citep{hernandez07b,hernandez08,hernandez10},
weak excess transition \citep{muzerolle10}.
Here we adopt the terminology ``evolved disk'' for this type
of object.
Some researchers include these objects in the transitional disk class.  However, these likely comprise a heterogenous
class of disks, including cleared inner disks, debris disks (see chapter in this volume by
{\it Matthews et al.}), and
disks with significant dust grain growth and settling.  This has been an issue in defining this subset of objects. We include evolved disks in this review for completeness, but focus on disks with more robust evidence for disk holes and gaps.

\begin{deluxetable}{lcccccc}
\tabletypesize{\small}
\tablecaption{Overview of Selected (Pre-)Transitional Disks}\label{tabbest}
\tablewidth{0pt}
\tablehead{
Object				& Class$^{a}$ & Submm Cavity & NIR Cavity & 0.1 M$_{\sun}$ Companion & Accretion Rate  &  References$^{b}$   \\
					&		& Radius (AU)  & detected?	& Detection Limit		& ($M_{\odot}\,yr^{-1}$) &
}
\startdata
AB Aur      		&	PTD &	70		& ... 		& 6 AU			& 1.3$\times$10$^{-7}$		&	1, 25, 38	\\ 
CoKu Tau$/$4		&	TD	&	...	& ... 		& binary$^{d}$	& $<$10$^{-10}$				& 	26, 39	\\
DM Tau				&	TD	&	19		& ... 		& 6 AU			& 2.9$\times$10$^{-9}$		& 	2, 25, 40	\\  
GM Aur				&	TD	&	28		& yes 			& 6 AU			& 9.6$\times$10$^{-9}$		& 	2, 11, 25, 40 	\\   
HD 100546			&	PTD	&	...	& yes 			& ...			& 5.9$\times$10$^{-8}$		& 	12, 41	\\ 
HD 141569			&	PTD	&	...	& yes 			& ...			& 7.4$\times$10$^{-9}$ 		&	13, 38	\\ 
HD 142527  	    	&	PTD	&	140		& yes	 		& binary$^{e}$	& 9.5$\times$10$^{-8}$ 		&	3, 14, 27, 38	\\ 
HD 169142			&	PTD	&	...	& yes 			& ...		& 9.1$\times$10$^{-9}$ 		&	15, 38	\\ 
IRS 48        		&	TD$^{c}$	&	60		& ... 		& 8 AU			& 4.0$\times$10$^{-9}$		&  	4, 28, 38	\\  
LkCa 15				&	PTD	&	50		& yes 			& 6 AU			& 3.1$\times$10$^{-9}$		& 	2, 16, 25, 40	\\   
MWC 758				&	PTD	&	73		& ... 		& 28 AU			& 4.5$\times$10$^{-8}$		& 	2, 29, 38	\\  
PDS 70				&	PTD	&	...	& yes 			& 6 AU			& $<$10$^{-10}$				& 	17, 30, 42	\\  
RX J1604-2130		&	TD	&	70  	& yes 			& 6 AU			& $<$10$^{-10}$ 			&	5, 18, 31	\\  
RX J1615-3255		&	TD	&	30		& no			& 8 AU			& 4$\times$10$^{-10}$ 		&	2, 19, 33, 43	\\ 
RX J1633-2442		&	TD	&	25  	& ... 		& 6 AU			& 1.3$\times$10$^{-10}$		& 	6, 25, 44	\\   
RY Tau    			&	PTD	&	14		& ... 		& 6 AU			& 6.4--9.1$\times$10$^{-8}$	&	7, 25, 45 	\\  
SAO 206462			&	PTD	&	46		& ... 		& 25 AU			& 4.5$\times$10$^{-9}$ 		&	2, 32, 38	\\	
SR 21				&	TD	&	36		& no			& 8 AU			& $<$1.4$\times$10$^{-9}$	& 	2, 20, 33, 46	\\	
SR 24 S				&	PTD	&	29		& ... 		& 8 AU			& 7.1$\times$10$^{-8}$		& 	2, 33, 46	\\
Sz 91				&	TD	&	65  	& yes 			& 25 AU			& 1.4$\times$10$^{-9}$ 		&	8, 21, 34	\\  
TW Hya				&	TD	&	4		& no 			& 3 AU			& 1.8$\times$10$^{-9}$		& 	9, 22, 35, 40 	\\ 
UX Tau A			&	PTD	&	25		& ... 		& 6 AU			& 1.1$\times$10$^{-9}$  	& 	2, 25, 40, 47 	\\ 	 
V4046 Sgr			&	TD	&	29  	& ... 		& binary$^{f}$	& 5.0$\times$10$^{-9}$ 		& 	10, 36, 48 	\\ 
DoAr 44				&	PTD	&	30		& no 			& 8 AU			& 3.7$\times$10$^{-9}$ 		& 	2, 23, 33, 38	\\ 
LkH$_{\alpha}$ 330	&	PTD	&	68		& no 			& 8 AU			& 2.2$\times$10$^{-8}$ 		& 	2, 24, 33, 38	\\ 
WSB 60				&	PTD	&	15		& ... 		& 25 AU			& 3.7$\times$10$^{-9}$ 		& 	2, 37, 46 	
\enddata
\vskip -.15 in
\tablenotetext{a} {We classify objects as either a pre-transitional disk (PTD) or transitional disk (TD).}
\tablenotetext{b}
{Submm cavity radii: 
(1) \citet{pietu05};
(2) \citet{andrews11};
(3) \citet{casassus12};
(4) \citet{bruderer14};
(5) \citet{mathews12};
(6) \citet{cieza12};
(7) \citet{isella10a};
(8) \citet{tsukagoshi13};
(9)	\citet{hughes07};
(10) \citet{rosenfeld13}.
NIR cavities: 
(11) {\em Hashimoto et al.}, in prep.;
(12) \citet{tatulli11};
(13) \citet{weinberger99};
(14) \citet{avenhaus14};
(15) \citet{quanz13b};
(16) \citet{thalmann10};
(17) \citet{hashimoto12};
(18) \citet{mayama12};
(19) {\em Kooistra et al.}, in prep.;  
(20) \citet{follette13}; 
(21) \citet{tsukagoshi13};
(22) {\em Akiyama et al.}, in prep.;
(23) {\em Kuzuhara et al.}, in prep.;
(24) {\em Bonnefoy et al.}, in prep.
Companion detection limits: 
(6);
(25) \citet{kraus11};
(26) \citet{ireland08};
(27) \citet{biller12};
(28) {\em Lacour et al.}, in prep.;
(29) \citet{grady13};
(30) {\em Kenworthy et al.}, in prep.;
(31) \citet{kraus08};
(32) \citet{vicente11};
(33) {\em Ireland et al.}, in prep.;
(34) \citet{romero12};
(35) \citet{evans12};
(36) \citet{stempels04};
(37) \citet{ratzka05}.
Accretion rates:
(5);
(38) \citet{salyk13};
(39) \citet{cohen79};
(40) \citet{ingleby13};
(41) \citet{pogodin12};
(42) \citet{dong12};
(43) \citet{krautter97};
(44) \citet{cieza10};
(45) \citet{calvet04};
(46) \citet{natta06};
(47) \citet{alcala14};
(48) \citet{donati11}.}
\tablenotetext{c} {The classification of IRS~48 as a TD or PTD is uncertain due the presence of strong PAH
emission in this object.}  
\tablenotetext{d} {The CoKu Tau$/$4 binary has a separation of 8~AU, which is  
within the range of semi-major axes required by tidal interaction theory to explain the 14~AU SED-inferred inner disk hole of this object \citep{nagel10}.
A binary system with an
eccentricity of 0.8 can clear out a region 3.5 times the semi-major axis
\citep[e.g.,][]{artymowicz94}. }
\tablenotetext{e} {The HD 142527 binary is separated by 13~AU. This
is not large enough to explain the disk gap.
\citet{casassus13} have disputed the presence of a companion, but recent results
from \citet{close14} may confirm it.}
\tablenotetext{f} {The V4046 Sgr binary has a separation of 0.045~AU, too small to clear out the disk hole.  \citet{rosenfeld13} provide an aperture masking companion limit that suggests there are no 0.1 M$_{\sun}$ companions down to 3 AU. }

\end{deluxetable}

\bigskip
\noindent
\textit{Model fitting}
\medskip

Detailed modeling of many of the above-mentioned SEDs has been
performed in order to infer the structure of these disks.
SEDs of transitional disks (i.e., objects with little or no NIR and MIR emission) have
been fit with models of inwardly truncated optically thick disks
\citep[e.g.,][]{rice03,calvet02}.  The inner edge or
``wall'' of the outer disk is frontally illuminated by the star,
dominating most of the emission seen in the IRS spectrum, particularly
 from $\sim$20--30~{\micron}.
Some of the holes in transitional disks are relatively
dust-free (e.g., DM Tau)  while  SED
model fitting indicates that others with strong 10~{\micron} silicate
emission have a small amount of optically thin dust within their disk holes
to explain this feature \citep[e.g., GM Aur;][]{calvet05}. 
Beyond $\sim$40~{\micron}, transitional disks have a contribution to
their SEDs from the outer disk. 
In pre\nobreakdashes-transitional disks, the observed SED can be fit with an optically thick inner disk separated by an
optically thin gap from an optically thick outer disk
\citep[e.g.,][]{brown07,espaillat07b,mulders10,dong12}. There is an inner wall located at the
dust sublimation radius which dominates the NIR (2--5~{\micron})
emission and should cast
a shadow on the outer disk (see \S~2.4).  
In a few cases, the optically thick inner disk of pre\nobreakdashes-transitional disks has been confirmed using 
NIR spectra \citep{espaillat10} following the methods of \citet{muzerolle03}.
Like the transitional disks, there is evidence for relatively dust-free gaps (e.g.,
UX Tau A) as well as gaps with some small,
optically thin dust to explain strong 10~{\micron} silicate emission
features \citep[e.g., LkCa 15;][]{espaillat07b}.  
The SEDs of evolved disks can be fit with
full disk models \citep[][]{sicilia11}, particularly in which the dust is
very settled towards the midplane \citep[e.g.,][]{espaillat12}.

\begin{figure}[ht!]
\begin{center}
\includegraphics[width=8.2cm]{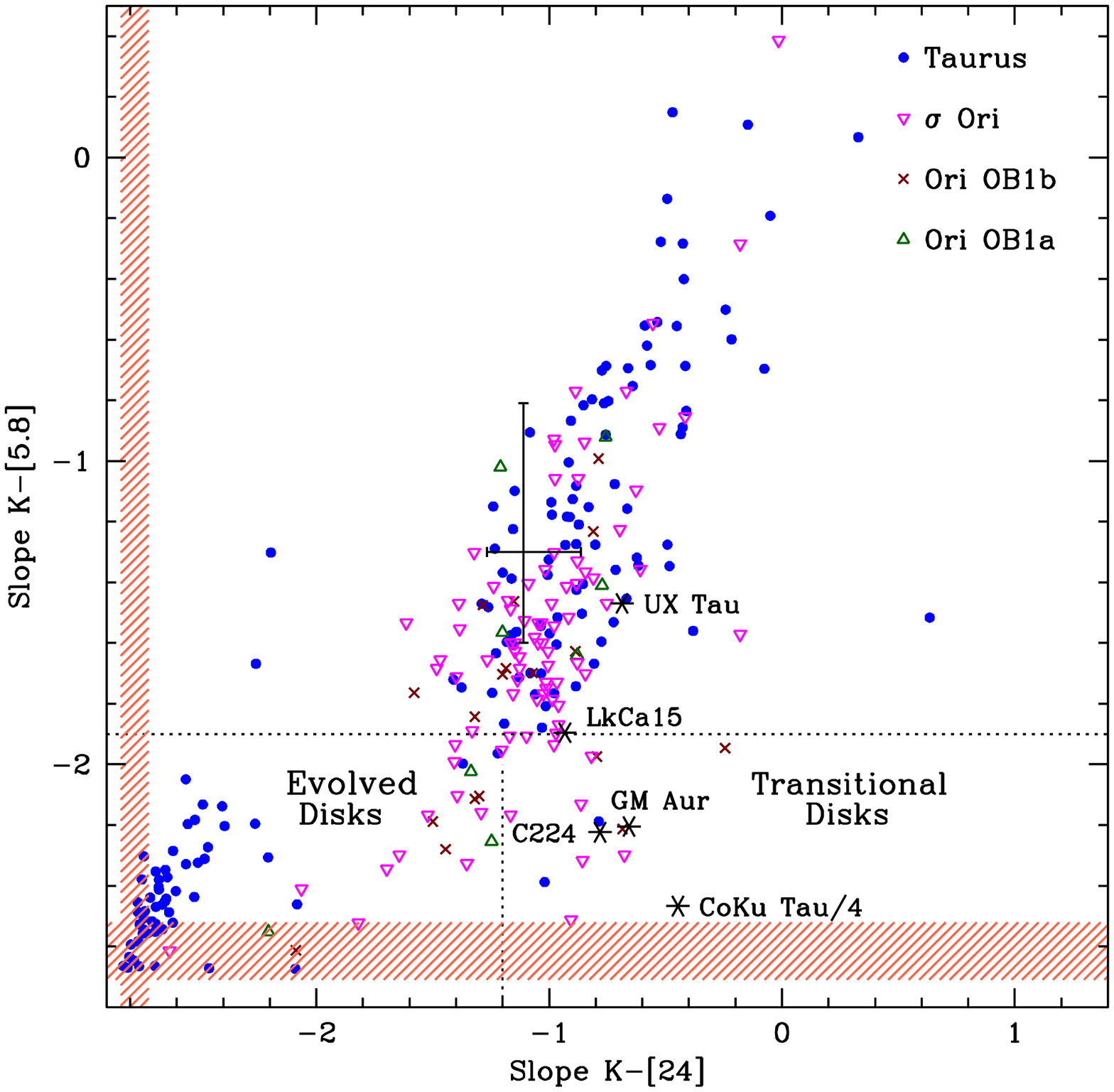}\\
\end{center}
 \vspace{-0.6 cm}
\caption{\small
{\it Spitzer} color-color diagram of objects
in different associations and clusters. 
Observations are shown for populations of different
ages: Taurus, 1-2 Myr \citep{luhman10}; 
$\sigma$ Ori, 3~Myr \citep{hernandez07a};
Ori OB1b, 7~Myr; Ori OB1a, 10~Myr \citep{hernandez07b}.
The hatched region corresponds to stellar photospheric colors.
The error bars represent the median and quartiles of Taurus objects
(i.e., where most full disks are expected to lie). 
Well characterized transitional 
disks \citep[GM Aur, CoKu Tau$/$4, CVSO224;][]{espaillat08b} and pre\nobreakdashes-transitional
disks (LkCa 15, UX Tau) are indicated with asterisks.
The dotted lines correspond to the lower quartile
of disk emission in $\sigma$ Ori, and roughly separate
the evolved disks (lower left) from the transitional disks (lower right).
Note that the pre\nobreakdashes-transitional disks do not lie below the dotted line, highlighting
that it is harder to identify disk gaps based on colors alone.
Figure adapted from \citet{hernandez07b}.
}
\label{figcolorcolor}
\end{figure}

There are many degeneracies to keep in mind when interpreting
SED-based results.
First, there is a limit to the gap sizes that
can be detected with {\it Spitzer} IRS.
Over 80$\%$ of the emission at 10~{\micron} comes from within 1~AU in the disk \citep[e.g.,][]{dalessio06}. Therefore, {\it Spitzer} IRS is most
sensitive to clearings in which a significant amount of dust located at radii
$<$1~AU has been removed, and so it will be easier to detect
disks with holes (i.e., transitional disks) as opposed to disks with gaps (i.e., pre\nobreakdashes-transitional disks).
The smallest gap in the innermost disk that will
cause a noticeable ``dip'' in the {\it Spitzer} spectrum would span $\sim$0.3-4~AU.  It would be
very difficult to detect gaps whose inner boundary is outside of 1~AU \citep[e.g.,~a gap
spanning  5--10~AU in the disk;][]{espaillat10}.  Therefore, with current data we cannot
exclude that any disk currently thought to be a full disk contains a small gap nor
can we exclude that currently known (pre\nobreakdashes-)transitional disks
have additional clearings at larger radii \citep[e.g.,][]{debes13}.  It will be largely up to
{\it ALMA} and the next generation of IR interferometers to detect such small disk gaps \citep[e.g.,][]{dejuanovelar13}.  
One should also keep in mind that
millimeter data are necessary to break the
degeneracy between dust settling and disk mass \citep[see][]{espaillat12}.
Also, the opacity of the disk is controlled by dust and in any sophisticated
disk model the largest uncertainty lies in the adopted dust opacities.
We will return to disk model limitations in \S~2.2.

\bigskip
\noindent
\textit{Implications for color-color diagrams}
\medskip

Another method of identifying transitional disks is through color-color diagrams (Fig.~\ref{figcolorcolor}).
This method grew in usage as more {\it Spitzer} IRAC and MIPS data became available and 
(pre-)transitional disks well characterized by {\it Spitzer} IRS spectra could be used to 
define the parameter space populated by these objects.
In these diagrams, transitional disks are 
distinct from other disks since they have NIR colors or slopes
(generally taken between two IRAC bands, or
K and an IRAC band)
significantly closer to stellar
photospheres than other disks in 
Taurus, but MIR colors (generally 
taken between K or one IRAC band and MIPS [24])
comparable or higher than other disks in Taurus
\citep[e.g.,][]{hernandez07a,hernandez08,hernandez10,merin10,muzerolle10,
luhman10,williams11,luhman12}.
Color-color diagrams are limited in their ability to 
identify pre\nobreakdashes-transitional disks because their fluxes in the NIR are comparable to many other disks in Taurus.
IRS data can do a better job of identifying pre\nobreakdashes-transitional disks
using the equivalent width of the 10~{\micron} feature or the NIR spectral index (e.g., n$_{2-6}$) versus the MIR spectral index \citep[e.g., n$_{13-31}$][]{furlan11,mcclure10,manoj11}. 
Evolved disks are easier to identify in color-color diagrams since they
show excesses over their stellar photospheres
that are consistently lower than most disks in Taurus, both in the near {\em and} mid-IR.

In the future, {\it JWST}'s sensitivity will allow us to expand {\it Spitzer}'s SED and color-color work to many more disks, particularly to fainter objects in older and farther star-forming regions, greatly increasing the known number of transitional, pre\nobreakdashes-transitional, and evolved disks.  Upcoming high-resolution imaging surveys with (sub)mm facilities in the near-future (i.e., ALMA) and IR interferometers further in the future (i.e., VLT$/$MATISSE) will give us a better understanding of the small-scale spatial structures in disks which SEDs cannot access.  

\begin{figure*}[ht!]
 \epsscale{2.1}
\plotone{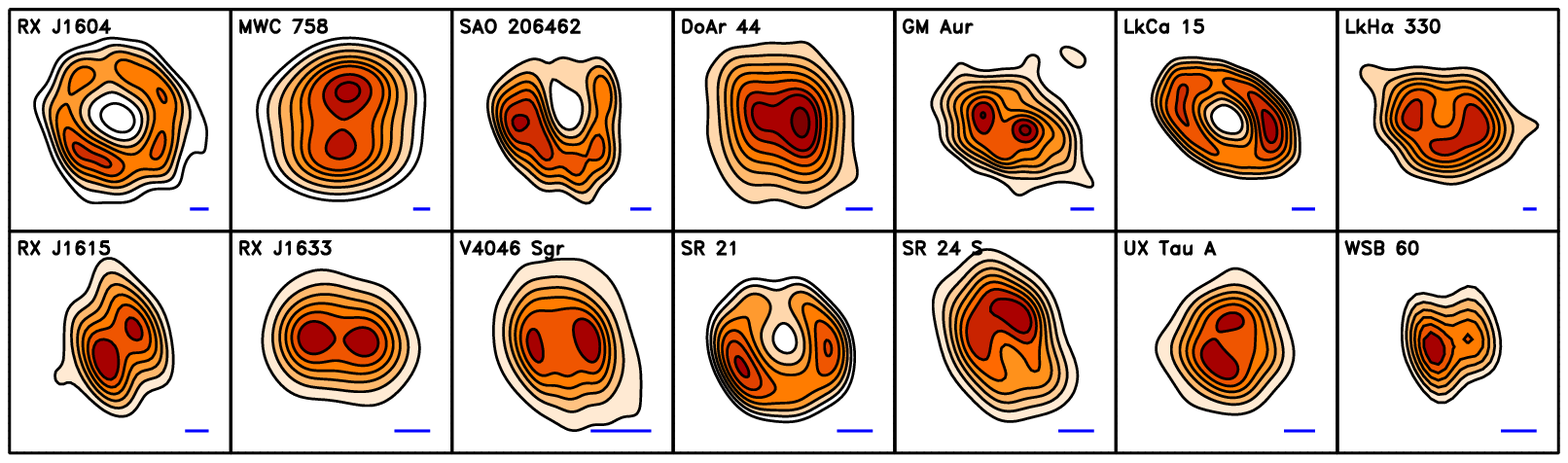}
 \vspace{-0.3 cm}
 \caption{\small 
A gallery of 880\,$\mu$m dust continuum images from the Submillimeter
Array for (pre\nobreakdashes-)transitional disks in nearby star-forming regions.  From left to right, 
the panels show the disks around RX J1604.3$-$2130 \citep{mathews12}, MWC 758 
\citep{isella10b}, SAO 206462 \citep{brown09}, DoAr 44 \citep{andrews09}, GM 
Aur \citep{hughes09}, LkCa 15 \citep{andrews11}, LkH$\alpha$ 330 
\citep{brown08}, RX J1615.3$-$3255 \citep{andrews11}, RX J1633.9$-$2442 
\citep{cieza12}, V4046 Sgr \citep{rosenfeld13}, SR 21 
\citep{brown09}, SR 24 S \citep{andrews10}, UX Tau A \citep{andrews11}, and WSB 
60 \citep{andrews09}.  A projected 30 AU scale bar is shown in blue at the 
lower right corner of each panel.
}  
\label{figmm}
\end{figure*}

\bigskip
\noindent
\textbf{ 2.2 Submillimeter$/$Radio Continuum Imaging}
\bigskip

The dust continuum at (sub)millimeter$/$radio wavelengths is an ideal probe of 
cool material in disks.  At these wavelengths, dust emission dominates 
over the contribution from the stellar photosphere, ensuring that contrast limitations are not an issue.  
Moreover, interferometers give access to the emission structure on a wide range 
of spatial scales, and will soon provide angular resolution that regularly 
exceeds 100\,mas.  The continuum emission at these 
long wavelengths is also thought to have relatively low optical depths, meaning 
the emission morphology is sensitive to the density distribution of mm and 
cm-sized grains \citep{beckwith90}.  These features are especially useful for observing 
the dust-depleted inner regions of (pre\nobreakdashes-)transitional disks, as will be
illustrated in the following subsections.  

\bigskip
\noindent
\textit{
(Sub)mm disk cavities} 
\medskip

With sufficient resolution, the (sub)mm dust emission from disks with cavities exhibits a 
``ring"-like morphology, with limb-brightened ansae along the major axis for 
projected viewing geometries.  In terms of the actual measured quantity, the 
interferometric visibilities, there is a distinctive oscillation pattern 
(effectively a Bessel function) where the first ``null" is a direct measure of 
the cavity dimensions \citep[see][]{hughes07}.  

As of this writing, roughly two dozen disk cavities have been 
directly resolved at (sub)mm wavelengths.  A gallery of representative 
continuum images, primarily from observations with the Submillimeter Array 
(SMA), is shown in Fig.~\ref{figmm}.  For the most part, these discoveries 
have been 
haphazard: some disks were specifically targeted based on their infrared SEDs
\citep[\S~2.1; e.g.,][]{brown09}, while others were found serendipitously in 
high resolution imaging surveys aimed at constraining the radial distributions 
of dust densities \citep[e.g.,][]{andrews09}.  

Perhaps the most remarkable aspect of these searches is the frequency of 
dust-depleted disk cavities at (sub)mm wavelengths, especially considering that 
the imaging census of all disks has so far been severely restricted by both 
sensitivity and resolution limitations.  In the nearest star-forming regions 
accessible to the northern hemisphere (Taurus and Ophiuchus), only about 
half of the disks in the bright half of the mm luminosity ($\sim$disk 
mass) distribution have been imaged with sufficient angular resolution 
($\sim$0.3$^{\prime\prime}$) to find large ($>$20\,AU in radius) disk cavities 
\citep{andrews09,andrews10,isella09,guilloteau11}.  
Even with these strong selection biases, the incidence of resolved 
cavities is surprisingly high compared to expectations from 
IR surveys (see \S~4). \citet{andrews11} estimated that at least 1 in 3 
of these mm-bright (massive) disks exhibit large cavities.  

\bigskip
\noindent
\textit{
Model fitting} 
\medskip

The basic structures of disk cavities can be quantified through radiative 
transfer modeling of their SEDs (see \S~2.1) simultaneously with resolved mm data.
These models often assume that 
the cavity can be described as a region of sharply reduced dust surface 
densities \citep[e.g.,][]{pietu06,brown08, hughes09, 
andrews11,cieza12,mathews12}. 
Such work finds cavity radii of $\sim$15--75\,AU, 
depletion factors in the inner disk of $\sim$10$^2$--10$^5$ relative to optically thick full disks, and outer regions with sizes and 
masses similar to those found for full disks.  However, there are subtleties 
in this simple modeling prescription.  First, 
the depletion levels are usually set by the infrared SED, not the mm data: 
the resolved images have a limited dynamic range and can only constrain an 
intensity drop by a factor $<$100.  Second, and related, is that the 
``sharpness" of the cavity edge is unclear.  The most popular model 
prescription implicitly imposes a discontinuity, but the data only directly 
indicate that the densities substantially decrease over a narrow radial range 
(a fraction of the still-coarse spatial resolution; $\sim$10\,AU).  Alternative 
models with a smoother taper at the cavity edge can explain the data equally 
well in many cases \citep[e.g.,][]{isella10b, isella12, andrews11b}, 
and might alleviate some of the tension with IR scattered 
light measurements (see \S~2.3).

Some additional problematic issues with these simple models have been 
illuminated, thanks to a new focus on the details of the resolved mm 
data.  For example, in some disks the dust ring morphology is found to be 
remarkably narrow -- with nearly all of the emission coming from a belt 
10-20\,AU across (or less) -- even as we trace gas with molecular line emission 
extending hundreds of AU beyond it \citep[e.g.,][]{rosenfeld13}.  This 
hints at the presence of a particle ``trap" near the cavity edge, as might be 
expected from local dynamical interactions between a planet and the gas disk 
\citep[see \S~3.2;][]{zhu12,pinilla12}.  In a perhaps related phenomenon 
\citep[e.g.,][]{regaly12,birnstiel13}, new 
high-fidelity images of (pre\nobreakdashes-)transitional disks are uncovering evidence that strong 
azimuthal asymmetries are common features of the mm emission rings 
\citep[][]{brown09,tang12,casassus13,vandermarel13,isella13}.  

\begin{figure}[ht]
\centering
 \includegraphics[width=8.2 cm]{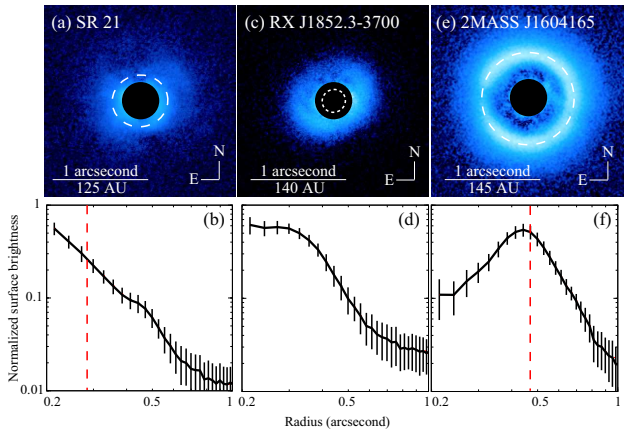}
 \vspace{-0.3 cm}
 \caption{\small
NIR polarimetric images of transitional disks in the $H$ band (top; 1.6~$\mu$m)
 along with their averaged radial surface brightness profiles along the major disk axis (bottom).
 Broken lines in the top and bottom panels correspond to
the radius of the outer wall as measured with submm imaging and$/$or SEDs.
We note that the regions of the inner disk that cannot be resolved are masked out in the panels.
From left to right the objects are as follows:  SR~21 \citep{follette13},
RX~J1852$-$3700 ({\em Kudo et al.}, in prep.),
RX~J1604$-$2130 \citep{mayama12}.
\label{figseeds}}
 \end{figure}

A number of pressing issues will soon be addressed by the ALMA project.  
Regarding the incidence of the disk holes and gaps, an expanded high 
resolution imaging census should determine the origin of the anomalously high 
occurrence of dust cavities in mm-wave images.  If the detection rate estimated by 
\citet{andrews11} is found to be valid at all luminosities, it would confirm 
that even the small amount of dust inside the disk cavities sometimes produces 
enough IR emission to hide the standard (pre\nobreakdashes-)transitional disk signature at short 
wavelengths, rendering IR selection inherently incomplete -- something already 
hinted at in the current data.  Perhaps more interesting would be evidence that the (pre\nobreakdashes-)transitional
disk frequency depends on environmental factors, like disk mass (i.e., a 
selection bias) or stellar host properties \citep[e.g.,][]{owen12b}.  
More detailed analyses of the disk structures are also necessary, both to 
develop a more appropriate modeling prescription and to better characterize the 
physical processes involved in clearing the disk cavities.  Specific efforts 
toward resolving the depletion zone at the cavity boundary, searching for 
material in the inner disk, determining ring widths and measuring their 
mm/radio colors to infer the signatures of particle evolution and trapping \citep[e.g.,][]{pinilla12,birnstiel13}, and quantifying 
the ring substructures should all impart substantial benefits on our 
understanding of disks.

\bigskip
\noindent
\textbf{ 2.3 Infrared Imaging}
\bigskip

IR imaging has been used successfully 
to observe disks around bright stars.
Space-based observations free from atmospheric turbulence
(e.g., HST) have detected fine disk structure
such as spiral features in the disk of HD~100546 \citep[][]{grady01}
and a ring-like gap in HD~141569 \citep[][]{weinberger99}.
MIR images of (pre-)transitional disks are able to trace
the irradiated outer wall, which effectively emits thermal radiation
\citep[e.g.,][]{geers07,maaskant13}.
More recently, high-resolution NIR polarimetric imaging and IR interferometry has become
available, allowing us to probe much further down into the inner  
few tens of AU in disks in nearby star-forming regions.
Here we focus on NIR polarimetric and IR interferometric imaging of
(pre\nobreakdashes-)transitional disks and also the results of IR imaging searches
for companions in these objects.

\bigskip
\noindent
\textit{NIR polarimetric imaging}
\medskip

NIR polarimetric imaging is capable of tracing the spatial distribution
of submicron-sized dust grains located in the uppermost, surface layer of disks.
One of the largest NIR polarimetry surveys of disks to date
was conducted as part of the SEEDS
\citep[Strategic Explorations of Exoplanets and Disks with Subaru;][]{tamura09}
project
\citep[e.g., Taurus at 140~pc;][{\it Tsukagoshi et al.}, submitted,] {thalmann10,tanii12,hashimoto12,mayama12,follette13,takami13}. 
There is also VLT$/$NACO imaging
work, predominantly focusing on disks around Herbig Ae$/$Be stars \citep[][]{quanz11,quanz12,quanz13a,quanz13b,rameau12}.
These surveys access the inner tens of AU in disks,
reaching a spatial resolution of 0.06$''$ (8 AU) in nearby star-forming regions.
Such observations have the potential to reveal fine structures such as 
spirals, warps, offsets, gaps, and dips in the disk. 

Most of the (pre\nobreakdashes-)transitional disks around TTS observed by SEEDS
have been resolved. This is because the stellar radiation can reach the outer disk more easily given that the innermost
regions of (pre\nobreakdashes-)transitional disks are less dense than full disks.
Many of these disks 
can be sorted into the three following categories
based on their observed scattered light emission (i.e.,  their ``polarized intensity'' appearance) at 1.6~$\mu$m: 

{\it (A) No cavity in the NIR with a smooth radial surface brightness profile
at the outer wall} (e.g, SR~21; DoAr~44; RX~J1615$-$3255; Fig.~\ref{figseeds}a, b),
 
{\it (B) Similar to category A, but with a broken radial brightness profile}
(e.g. TW~Hya; RX~J1852.3$-$3700; LkH$\alpha$~330). These disks
display a slight slope in the radial brightness profile in the inner portion of the disk, but a steep slope in the 
outer regions (Fig.~\ref{figseeds}c, d),

{\it (C) A clear cavity in the NIR polarized light}
(e.g., GM~Aur; Sz~91; PDS~70; RX~J1604-2130; 
Fig.~\ref{figseeds}e, f). \\

The above categories demonstrate that the spatial distribution of small and large dust grains in 
the disk are not necessarily similar.  
Based on previous submm images (see \S~2.2), the large, mm-sized dust grains in the inner
regions of each 
of the above disks is significantly depleted. However, in categories A and B
there is evidence that a significant amount of small, submicron sized dust grains remains in the inner disk,
well within the cavity seen in the submm images.  In category C, the small dust grains appear
to more closely trace the large dust distribution, as both are significantly depleted 
in the inner disk.  
One possible mechanism that could explain the differences between
the three categories presented above is dust filtration \citep[e.g.,][]{rice06,zhu12},
which we will return to in more detail in \S~3.2. 
More high-resolution imaging observations of disks at
different wavelengths is necessary to develop a fuller picture of their structure
given that the disk's appearance at a
certain wavelength depends on the dust opacity.

\bigskip
\noindent
\textit{IR interferometric imaging}
\medskip

IR interferometers, such as the Very Large Telescope Interferometer (VLTI), 
the Keck Interferometer (KI), and the CHARA array, 
provide milliarcsecond angular resolution
in the NIR and MIR regime (1--13\,$\mu$m),
enabling new constraints
on the structure of (pre\nobreakdashes-)transitional disks.
Such spatially resolved studies are important
to reveal complex structure in the small dust distribution within the innermost region
of the disk, testing the basic constructs of models that have been derived
based on spatially unresolved data (e.g., SEDs).

The visibility amplitudes measured with interferometry
permit direct constraints on the brightness profile,
and, through radiative transfer modeling (see \S~2.1 and \S~2.2 for
discussion of limitations), on the distribution 
and physical conditions of the circumstellar material.
The NIR emission ($H$ and $K$ band, $1.4-2.5\,\mu$m) in
the pre\nobreakdashes-transitional disks studied most extensively
with IR interferometry (i.e., HD\,100546, T\,Cha, and V1247\,Ori) 
is dominated by 
hot optically thick dust, with smaller contributions 
from scattered light \citep{olofsson13} and optically thin
dust emission \citep{kraus13}.
The measured inner disk radii are in general consistent with 
the expected location of the dust sublimation radius,
while the radial extent of this inner emission component
varies significantly for different sources
(HD\,100546: 0.24--4\,AU, \citealt{benisty10,tatulli11};
T\,Cha: 0.07--0.11\,AU, \citealt{olofsson11,olofsson13};
V1247\,Ori: 0.18--0.27\,AU, \citealt{kraus13}).

The MIR regime ($N$ band, $8-13\,\mu$m) is sensitive to 
a wider range of dust temperatures and stellocentric radii.
In the transitional disk of TW\,Hya, the region inside of
$0.3-0.52$\,AU is found to contain only optically thin dust 
\citep{eisner06,ratzka07,akeson11}, 
followed by an optically thick outer disk \citep{arnold12twhya}, in agreement
with SED modeling \citep{calvet02} and (sub)mm imaging
\citep{hughes07}. 
The gaps in the pre\nobreakdashes-transitional disks of the Herbig Ae$/$Be star HD\,100546 \citep{mulders13} and the TTS T\,Cha \citep{olofsson11,olofsson13} were found to be
highly depleted of (sub)$\mu$m-sized dust grains,
with no significant NIR or MIR emission,
consistent with SED-based expectations (i.e., no substantial 10~{\micron} silicate
emission).
The disk around the Herbig Ae$/$Be star V1247\,Ori, on the other hand, exhibits a gap
filled with optically thin dust.
The presence of such optically thin material within the gap is
not evident from the SED, while the interferometric
observations indicate that this gap material is
the dominant contributor at MIR wavelengths.
This illustrates the importance of IR interferometry 
for unraveling the physical conditions in disk gaps and holes.

We note that besides the dust continuum emission, some (pre\nobreakdashes-)transitional
disks exhibit polycyclic aromatic hydrocarbon (PAH)
spectral features, for instance at 7.7\,$\mu$m, 8.6\,$\mu$m, 
and 11.3\,$\mu$m.  For a few objects, it
was possible to locate the spatial origin of these features
using adaptive optics imaging \citep{habart06} or 
MIDI long-baseline interferometry \citep{kraus13b}.
These observations showed that these molecular bands
originate from a significantly more extended region than the 
NIR$/$MIR continuum emission, including the gap region and
the outer disk.  This is consistent with the 
scenario that these particles are transiently heated
by UV photons and can be observed over a wide range of
stellocentric radii.

One of the most intriguing findings obtained with IR
interferometry is the
detection of non-zero phase signals, which indicate
the presence of significant asymmetries in the 
inner, AU-scale disk regions.
Keck/NIRC2 aperture masking observations of V1247\,Ori \citep{kraus13}
revealed asymmetries whose
direction is not aligned with the disk minor axis
and also changes with wavelength.
Therefore, these asymmetries are neither consistent with 
a companion detection, nor with disk 
features.  Instead, these observations suggest the 
presence of complex, radially extended disk structures,
located within the gap region.  
It is possible that these structures are related 
to the spiral-like inhomogeneities that have been detected 
with coronagraphic imaging on about 10-times larger scales
\citep[e.g.,][]{hashimoto11,grady13} and that they reflect the dynamical interaction 
of the gap-opening body$/$bodies with the disk material.
Studying these complex density structures and relating the asymmetries
to the known spectro-photometric variability of these objects 
(\S~2.4) will be a major objective of future interferometric 
imaging studies.

The major limitations from the existing studies
arise from sparse $uv$-coverage, which has so far
prevented the reconstruction of direct interferometric images
for these objects.
Different strategies have been employed in order to
relax the $uv$-coverage restrictions, including the 
combination of long-baseline interferometric 
data with single-aperture interferometry techniques
(e.g.,\ speckle and aperture masking interferometry;
\citealt{arnold12twhya,olofsson13,kraus13}) and the combination of data from
different facilities \citep{akeson11,olofsson13,kraus13}.
Truly transformational results can be expected from the
upcoming generation of imaging-optimized long-baseline 
interferometric instruments, such as the 4-telescope 
MIR beam combiner MATISSE, which will enable 
efficient long-baseline interferometric imaging on scales of several AU.

\bigskip
\noindent
\textit{Companion Detections}
\medskip

NIR imaging observations can directly reveal companions within the cleared regions of disks. Both theory and observations have long shown that stellar binary companions can open gaps \citep[e.g.,][]{artymowicz94,jensen97,white01}, based on numerous moderate-contrast companions ($\sim$0--3 magnitudes of contrast, or companion masses $> 0.1 M_{\odot}$) that have been identified with RV monitoring, HST imaging, adaptive optics imaging, and speckle interferometry. 
For example, NIR imaging of CoKu Tau$/$4, a star surrounded by a transitional disk, revealed a previously unknown stellar-mass companion that is likely responsible for the inner clearing in this disk, demonstrating that
it is very important to survey stars with (pre\nobreakdashes-)transitional disks for binarity in addition to exploring other possible clearing mechanisms \citep[see \S~3.2;][]{ireland08}.  

The detection of substellar or planetary companions has been more challenging, due to the need for high contrast ($\Delta L' = $5 to achieve $M_{lim} = 30 M_{Jup}$ for a 1 $M_{\odot}$ primary star) near or inside the formal diffraction limit of large telescopes. Most of the high-contrast candidate companions identified to date have been observed with interferometric techniques such as nonredundant mask interferometry \citep[NRM;][]{tuthill00,ireland12}, which measure more stable observable quantities (such as closure phase) to achieve limits of $\Delta L' = $7--8 at $\lambda /D$ (3--5 $M_{Jup}$ at 8 AU). The discoveries of NRM include a candidate planetary-mass companion to LkCa 15 \citep{kraus12} and a candidate low-mass stellar companion to T Cha \citep{huelamo11}.  A possible candidate companion was reported around FL Cha, although these asymmetries could be associated with disk emission instead \citep{cieza13}.

Advanced imaging techniques also are beginning to reveal candidate companions at intermediate orbital radii that correspond to the optically thick outer regions of (pre-)transitional disks \citep[e.g.,][]{quanz13a}, beyond the outer edge of the hole or gap region. The flux contributions of companions can be difficult to distinguish from scattered light due to disk features \citep[][]{olofsson13,cieza13,kraus13}. However, the case of LkCa 15 shows that the planetary hypothesis can be tested using multi-epoch, multi-wavelength data (to confirm colors and Keplerian orbital motion) and by direct comparison to resolved submm maps (to localize the candidate companion with respect to the inner disk edge).

Even with the enhanced resolution and contrast of techniques like NRM, current surveys are only able to probe super-Jupiter masses in outer solar systems. For bright stars ($I \la 9$), upcoming extreme adaptive optics systems like GPI and SPHERE will pave the way to higher contrasts with both imaging and NRM, achieving contrasts of $\Delta K \ge 10$ at $\lambda /D$ ($\sim$1--2 $M_{Jup}$ at 10 AU). However, most young solar-type and low-mass stars fall below the optical flux limits of extreme AO. Further advances for those targets will require observations with JWST that probe the sub-Jupiter regime for outer solar systems ($\Delta M \sim$10 at $\lambda /D$, or $<$1 $M_{Jup}$ at $>$15 AU) or with future ground-based telescopes that probe the Jupiter regime near the snow line (achieving $\Delta K \sim$10 at $\lambda /D$ or $\sim$1--2 $M_{Jup}$ at $>$2--3 AU).\\

\bigskip
\noindent
\textbf{ 2.4 Time domain studies}
\bigskip

IR variability in TTS is ubiquitous and several ground-based studies
have been undertaken to ascertain the nature of this variability
\citep[e.g.,][]{joy45,rydgren76,carpenter01}.
With the simultaneous MIR wavelength coverage provided by
{\it Spitzer} IRS, striking variability in (pre\nobreakdashes)-transitional disks was discovered,
suggestive of structural changes in these disks with time.  We review this variability along
with the mechanisms that have been proposed to be responsible for it in the following subsections.

\bigskip
\noindent
\textit{
``Seesaw'' variability} 
\medskip

\begin{figure}[htb]
\begin{center}
 \includegraphics[width=8.2cm]{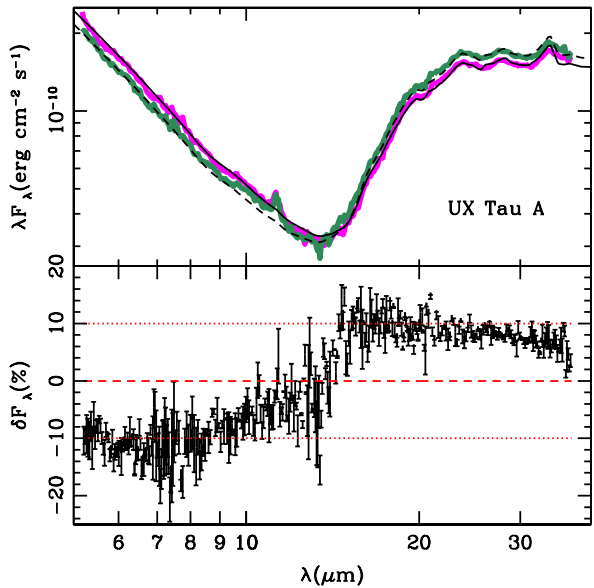}
  \end{center}
   \vspace{-0.3 cm}
 \caption{\small Top: Observed {\it Spitzer} IRS spectra (gray) and models (solid and broken black lines) for the pre\nobreakdashes-transitional disk of UX~Tau~A.  The variability between the two spectra can be reproduced by changing the height of the inner disk wall in the model by 17$\%$.   Bottom: Percentage change in flux between the two IRS spectra above.  The observed variability cannot be explained by the observational uncertainties of IRS (error bars).  Figure adapted from \citet{espaillat11}.
}
\label{figvar1}
 \end{figure}

The flux in many pre\nobreakdashes-transitional disks observed for variability to date ``seesaws,'' i.e., as the emission decreases at
shorter wavelengths in the IRS data, the emission increases at
longer wavelengths \citep[Fig.~\ref{figvar1};][]{muzerolle09,espaillat11,flaherty12}.
MIR variability with IRS was also seen in some transitional disks \citep[e.g., GM~Aur and LRLL~67;][]{espaillat11,flaherty12}, though in these objects the variability was predominantly around the
region of the silicate emission feature.
Typically, the flux in the pre\nobreakdashes-transitional disks and transitional disks observed changed by about 10$\%$ between epochs, but in some objects the change was as high as 50$\%$.

\begin{figure}[htb]
\begin{center}
 \includegraphics[width=8.4cm]{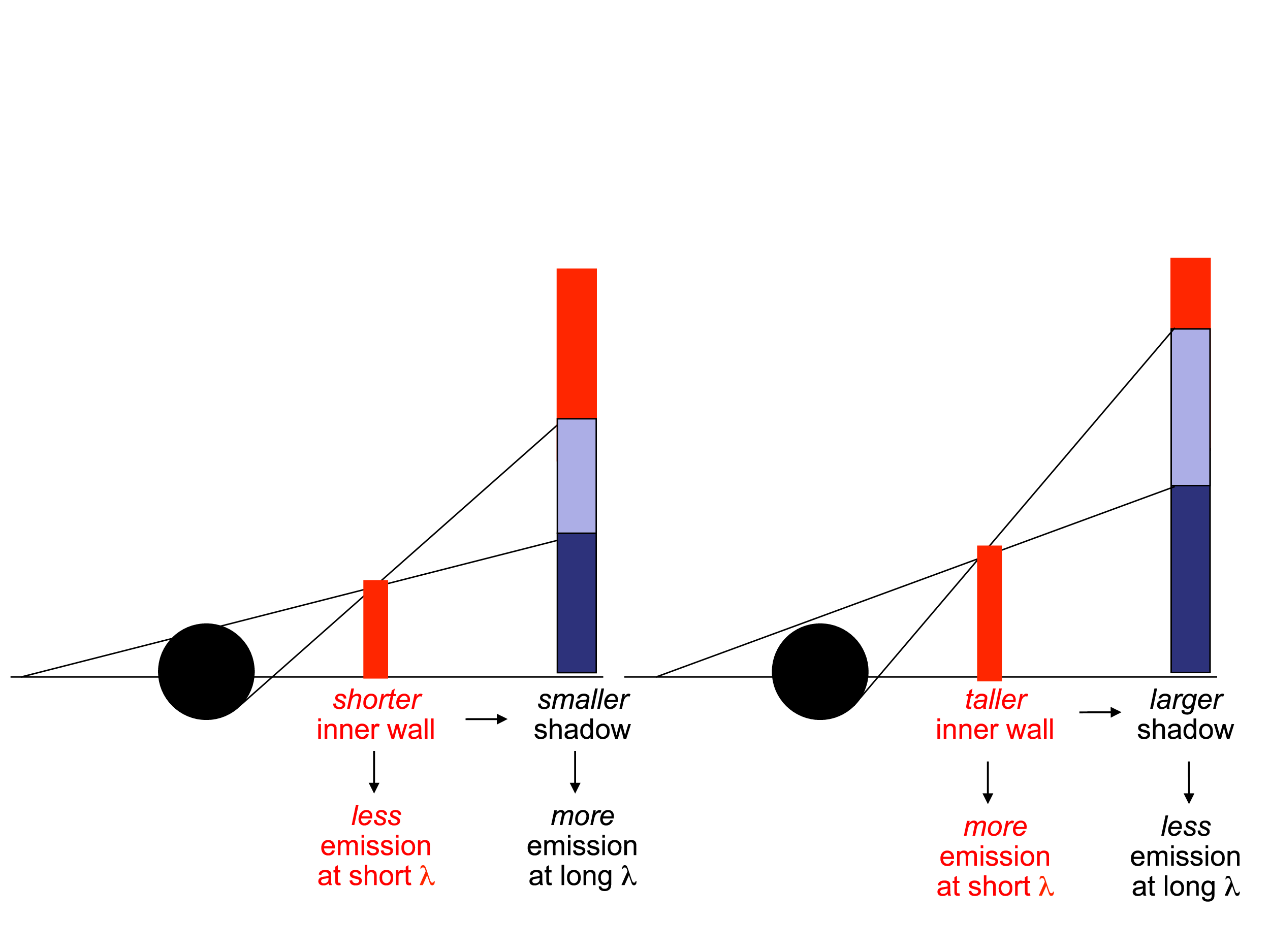}
  \end{center}
   \vspace{-0.3 cm}
 \caption{\small Schematic depicting the proposed link between the observed variability and disk structure based on the SED modeling from Fig.~\ref{figvar1}.  Red corresponds to visible areas of the disk wall, while light blue and dark blue areas are in the penumbra and umbra, respectively. The MIR variability observed in pre\nobreakdashes-transitional disks can be explained by changes in the height of the inner disk wall, which results in variable shadowing of the outer wall.  
}
\label{figvar2}
 \end{figure}

This variability may point to structural changes in disks. 
SED modeling can explain the seesaw behavior seen in pre\nobreakdashes-transitional disks by changing the height of the inner wall of these disks \citep[Fig.~\ref{figvar2};][]{espaillat11}.  When the inner wall is taller, the emission at the shorter wavelengths is higher since the inner wall dominates the emission at 2-8 ~{\micron}. The taller inner wall casts a larger shadow on the outer disk wall, leading to less emission at wavelengths beyond 20 ~{\micron} where the outer wall dominates. When the inner wall is shorter, the emission at the shorter wavelengths is lower and the shorter inner wall casts a smaller shadow on the outer disk wall, leading to more emission at longer wavelengths. This ``seesaw'' variability confirms the presence of optically thick material in the inner disk of pre\nobreakdashes-transitional disks.
The variability seen in transitional disks may suggest that while the disk is vertically optically thin, there is a radially optically thick structure in the inner disk, perhaps composed of large grains and$/$or limited in spatial extent 
so that it does not contribute substantially to the emission between 1--5~{\micron} while still leading to
shadowing of the outer disk.  One intriguing possibility involves accretion streams connecting multiple planets, as predicted in the models of \citet{zhu11, dodson11} and claimed
to be seen by ALMA \citep{casassus13}.  

\bigskip
\noindent
\textit{
Possible underlying mechanisms} 
\medskip

When comparing the nature of the observed variability to currently known physical mechanisms, it seems 
unlikely that star spots, winds, and stellar magnetic fields are the underlying
cause.
The star spots proposed to explain
variability at shorter wavelengths in other works \citep[e.g.,][]{carpenter01} could change the irradiation heating, but
this would cause an overall increase or decrease of the flux, not seesaw variability.
A disk wind which carries dust may shadow the outer disk.  
However, \citet{flaherty11} do not find evidence for strong winds in their sample which displays
seesaw variability.
Stellar magnetic fields that interact with dust beyond the dust sublimation radius may lead to changes
if the field expands and contracts or is tilted with respect to plane of the disk
\citep{goodson99,lai08}.
However, it is thought that the stellar magnetic field truncates the disk within the corotation radius
and for many objects.  The corotation radius is within the dust sublimation radius, making it unlikely
that the stellar magnetic field is interacting with the dust in the disk \citep{flaherty11}.

It is unclear what role accretion or X-ray flares may play in disk variability.  Accretion rates are known to
be variable in young objects, but \citet{flaherty12} do not find that the observed variations in accretion rate are
large enough to reproduce the magnitude of the MIR variability observed.
Strong X-ray flares can increase the ionization of
dust and lead to a change in scale height \citep{ke12}.
However, while TTS are known to have strong X-ray flares \citep{feigelson07},  
it is unlikely that all of the MIR disk variability observed overlapped with strong 
X-ray flares.

The MIR variability seen is most likely due to perturbations in the disk, possibly by planets or turbulence.
Planets are thought to create spiral density waves in the disk \citep[see PPV review by][]{durisen07},
which may have already been detected in disks \citep[e.g.,][]{hashimoto11,grady13}.
Such spiral density waves may affect the innermost disk, 
causing the height of the inner disk wall to change
with time and creating the seesaw variability observed.   
The timescales of the variability discussed here span $\sim$1--3~years down to 1~week or less.  If the variability is related to an orbital timescale, this corresponds to $\sim$1--2~AU and $<$0.07~AU in the disk, plausible locations for planetary companions 
given our own solar system and detections of hot Jupiters \citep{marcy05}.
Turbulence is also a viable solution.   Magnetic fields in a turbulent disk may lift dust and gas off the disk 
\citep{turner10,hirose11}.  The predicted magnitude of such
changes in the disk scale height are consistent with the observations.

Observations with {\it JWST} can explore the range of timescales and
variability present in (pre\nobreakdashes-)transitional disks to test all of the scenarios explored above.
It is also likely that a diverse range of variability is present in most disks around young stars \citep[e.g., YSOVAR;][]{morales11} and
{\it JWST} observations of a wide range of disks will help us more fully categorize
disk variability.

\bigskip
\noindent
\textbf{ 2.5 Gaseous Emission}
\bigskip

\begin{figure}[htb]
\begin{center}
 \includegraphics[width=8.2cm]{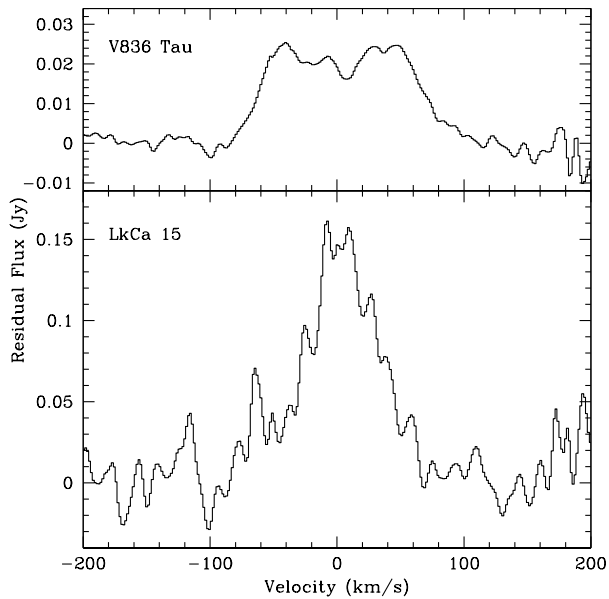}
 \end{center}
  \vspace{-0.3 cm}
 \caption{\small Rovibrational line profiles of CO $v$=1--0 emission 
from V836 Tau (top)  and LkCa~15 (bottom)  after correction for 
stellar CO absorption \citep{najita08}.  The double-peaked 
profile of V836~Tau indicates that the gas emission is truncated 
beyond $\sim 0.4~$\,AU.  
The centrally-peaked LkCa~15 profile indicates gas emission extending to 
much larger radii. 
}
\label{figco}
 \end{figure}

The structure of the gas in (pre\nobreakdashes-)transitional disks is a valuable probe,  
because the mechanisms proposed to account for the properties of 
these objects
(\S~\ref{sec:theory}; e.g., grain growth, photoevaporation, companions) may impact the gas in different ways from the dust.
The chapters in this volume by {\it Pontoppidan et al.}\ and {\it Alexander et al.}\, 
as well as the earlier PPV review by \citet{najita07b}, 
describe some of the available atomic and molecular diagnostics 
and how they are used to study gas disks.  
In selecting among these specifically for the purpose of probing radial 
disk structure, it is important to consider how much gas 
is expected to remain in a hole or gap. 
Although theoretical studies suggest the gas column density is 
significantly reduced in the cleared dust region \citep[e.g., by $\sim 1000$;][]{regaly10},  
given a typical TTS's disk column density of 100\,g/cm$^{2}$ at 1\,AU,   
a fairly hefty gas column density could remain. 
Many stars hosting (pre\nobreakdashes-)transitional disks also show signs of significant
gas accretion (\S~4.2).
These considerations suggest that molecular diagnostics, which probe 
larger disk column densities ($N_H>$ 1$\times$10$^{21}$ cm$^{-2}$), 
are more likely to be successful in detecting a hole or gap in the gas disk.
In the following, we review what is known to date about the radial structure
of gas in disks.

\bigskip
\noindent
\textit{
MIR and (sub)mm spectral lines}
\medskip

As there are now gas diagnostics that probe disks 
over a wide range of radii ($\sim 0.1-100$\,AU), 
the presence or absence of these diagnostics in emission can 
give a rough idea 
of whether gas disks are radially continuous or not. 
At large radii, 
the outer disks of many (pre\nobreakdashes-)transitional disks
(TW Hya, GM Aur, DM Tau, LkCa~15, etc.) 
are well studied at mm wavelengths 
\citep[e.g.,][see also \S~2.2]{koerner95,guilloteau94,dutrey08}.
For some, the mm observations indicate that 
rotationally excited CO exists within the cavity in the dust distribution
\citep[e.g., TW Hya, LkCa~15;][]{rosenfeld12,pietu07},
although these observations cannot currently constrain how that gas 
is distributed. 
Many (pre\nobreakdashes-)transitional disks also 
show multiple signatures of gas close to the star: 
ongoing accretion (\S~4.2), 
UV H$_{2}$ emission \citep[e.g.,][]{ingleby09,france12}, 
and rovibrational CO emission 
\citep{salyk09,salyk11b,najita08,najita09}.  

In contrast, the 10-20~$\mu$m {\it Spitzer} spectra of these objects
conspicuously lack the rich
molecular emission (e.g., H$_{2}$O, C$_{2}$H$_{2}$, HCN) that characterizes
the mid-infrared spectra of full disks around classical TTS
\citep[CTTS, i.e., stars that are accreting;][]{najita10, pontoppidan10}. 
As these MIR molecular diagnostics probe radii within a few AU of the star, 
their absence is suggestive of a missing molecular disk at these radii, 
i.e., a gap between an inner disk (traced by CO and UV H$_{2}$) and 
the outer disk (probed in the mm). 
Alternatively, the disk might be too cool to emit at these 
radii, or the gas may be abundant but in atomic form. 
Further work, theoretical and$/$or observational, is needed to 
evaluate these possibilities.

\bigskip
\noindent
\textit{
Velocity resolved spectroscopy} 
\medskip

Several approaches can be used to probe the distribution of the
gas in greater detail.
In the absence of spatially resolved imaging, which is the most 
robust approach, 
velocity resolved spectroscopy
coupled with the assumption of Keplerian rotation can probe the 
radial structure of gaseous disks 
\citep[see PPIV review by][]{najita00}.
The addition of spectroastrometric information (i.e., the spatial centroid 
of spectrally resolved line emission as a function of velocity) can reduce  
ambiguities in the disk properties inferred with this approach 
\citep{pontoppidan08}.
These techniques have been used to search for sharp changes in gaseous 
emission as a function of radius as evidence of disk cavities, 
and to identify departures from azimuthal symmetry such as those 
created by orbiting companions.

Velocity resolved spectroscopy of CO rovibrational emission provides 
tentative evidence for a truncated inner gas disk in 
the evolved disk V836 Tau \citep{strom89}.
An optically thin gap, if there is one, would be found at small radii ($\sim 1$\,AU) 
and plausibly overlap the range of disk radii probed by the CO emission. 
Indeed, the CO emission from V836 Tau is unusual in showing a 
distinct double-peaked profile consistent with the truncation of the 
CO emission beyond $\sim 0.4$\,AU \citep[Fig.~\ref{figco};][]{najita08}. 
In comparison, other disks show much more 
centrally peaked rovibrational CO line profiles 
\citep{salyk07,salyk09,najita08,najita09}.

\begin{figure}[htb]
\begin{center}
 \includegraphics[width=7.2cm]{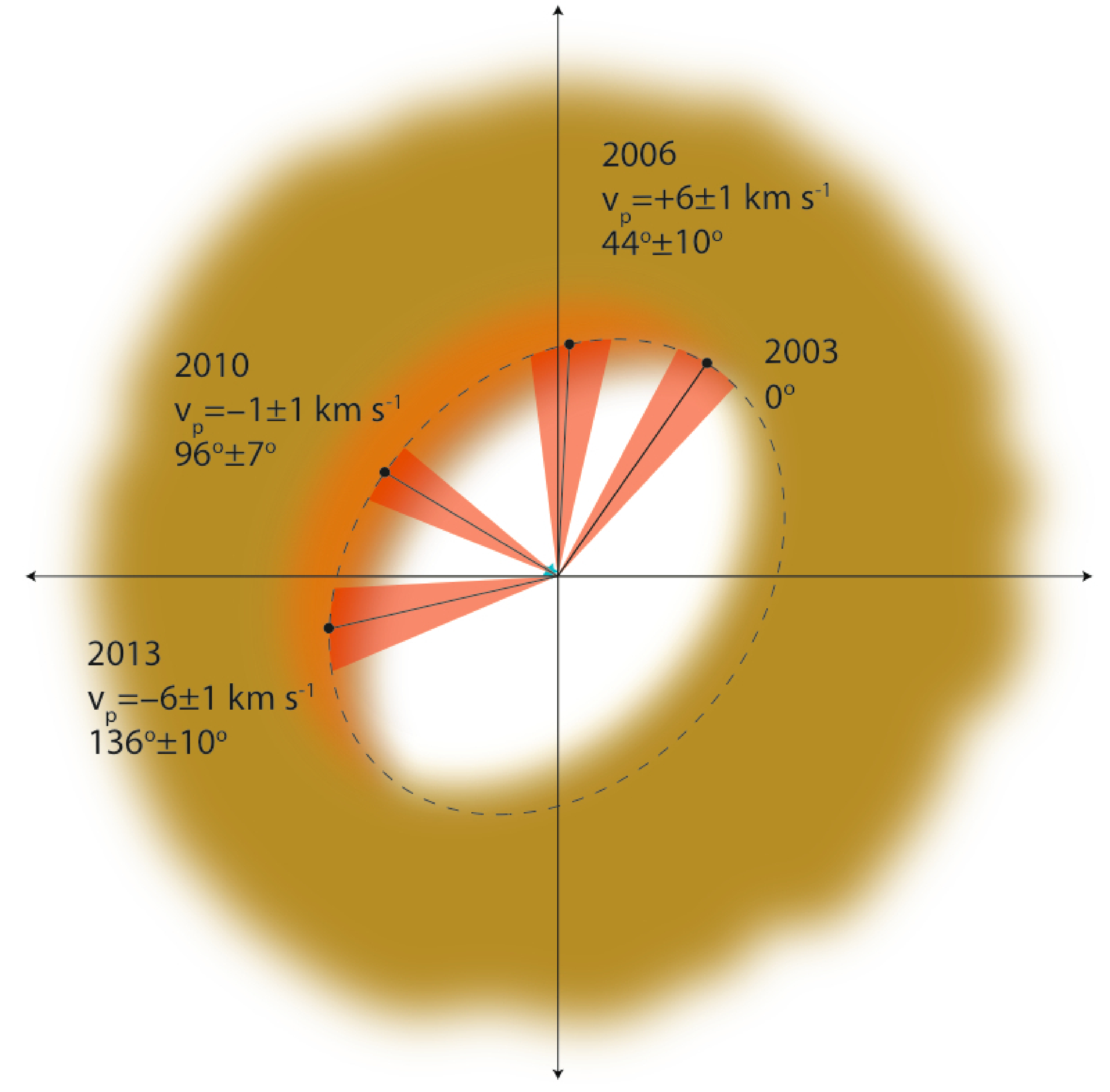}
  \end{center}
   \vspace{-0.3 cm}
 \caption{\small Spectroastrometry of the CO rovibrational emission from HD100546,
followed over 10 years, reveals evidence for a compact source of CO
emission that orbits the star near the inner disk edge \citep[][and 2014, in
preparation]{brittain13}. Possible interpretations include emission from a circumplanetary 
disk or a hot spot heated by an orbiting companion.  
}
\label{figoh}
 \end{figure}

Spectroscopy also indicates possible differences in the radial extent 
of the gas and dust in the inner disk.
The CO emission profile from the pre\nobreakdashes-transitional disk of LkCa~15 
\citep[Fig.~\ref{figco};][]{najita08} 
spans a broad range of velocities,  
indicating that the inner gas disk extends over a much larger range of 
radii (from $\sim 0.08$\,AU out to several AU) than in V836~Tau. 
This result might be surprising given that SED modeling suggests 
that the inner optically thick dust disk of LkCa~15 
extends over a narrow annular region 
\citep[0.15--0.19\,AU;][]{espaillat10}. 
The origin of possible differences in the radial extent of the gas and dust in 
the inner disk region is an interesting topic for future work. 

Some of the best evidence to date for the truncation of the 
outer gas disks of (pre\nobreakdashes-)transitional disks comes from studies of Herbig Ae$/$Be stars. 
For nearby systems with large dust cavities, 
ground-based observations can 
spatially resolve the inner edge of the CO rovibrational emission 
from the outer disks around these bright stars \citep[e.g., HD141569;][]{goto06}. 
Line profile shapes and constraints from UV fluorescence modeling have also been used to show that 
rovibrational CO and OH emission is truncated at the same 
radius as the disk continuum in some systems
\citep[][]{brittain07,vanderplas09,brittain09,liskowsky12},
although the radial distribution of the gas and dust appear to differ 
in other systems \citep[e.g., IRS 48;][]{brown12b}. 
The gas disk may sometimes be truncated significantly inward of the outer dust disk and far from the star. 
For example, the CO emission from SR 21 extends much further in (to $\sim$7~AU) than the inner hole size of $\sim$18~AU 
\citep{pontoppidan08}.

Several unusual aspects of the gaseous emission from HD100546
point to the possibility that the cavity in the molecular
emission is created by an orbiting high mass giant planet.
\cite{acke06} inferred the presence of a gap in the gas disk based on a local minimum in the  [\ion{O}{1}] 6300~{\AA} emission from the disk. 
The rovibrational OH emission from HD100546 is found to show a strong line
asymmetry that is consistent with emission from an
eccentric inner rim \citep[$e \ge 0.18$;][]{liskowsky12}.
Eccentricities of that magnitude are predicted to result from
planet-disk interactions \citep[e.g.,][]{papaloizou01, kley06}.
In addition, the CO rovibrational emission varies in its
spectroastrometric signal. 
The variations can be explained by a CO emission component that
orbits the star just within the 13\,AU inner rim \citep[Fig.~\ref{figoh};][]{brittain13}.  
The required emitting area (0.1\,AU$^2$) is similar to that 
expected for a circumplanetary disk surrounding a $5 M_J$ planet 
at that distance \citep[e.g.,][]{quillen98,martin11}. 
Further studies with ALMA and ELTs will give us 
new opportunities to explore the nature of disk holes and gaps 
in the gas disk 
\citep[e.g.,][]{vandermarel13, casassus13}.

\bigskip
\section{\textbf{COMPARISON TO THEORETICAL MECHANISMS}} \label{sec:theory}
\bigskip

There are three main observational constraints every theory must confront when attempting
to explain (pre\nobreakdashes-)transitional disk observations.
First, the cleared regions in (pre\nobreakdashes-)transitional disks studied
to date are generally large (\S~2.1, \S~2.2). For transitional disks,
the optically thin region extends from tens of AU all the way down to the central star.
For pre\nobreakdashes-transitional disks, SED modeling
suggests that the optically thick region may only extend up to $\sim$ 1 AU,
followed by a gap that is significantly depleted of dust up to tens of AU, as seen in the transitional disks.
At the same time, submm imaging has revealed the existence
of some disks which have inner regions significantly depleted of large dust grains without
exhibiting MIR SED deficits (\S~2.2), suggesting that a large amount of small dust
grains still remain in the inner disk, as also shown by NIR polarimetric images (\S~2.3).
Second, in order to be discernible in the SED, the gaps$/$holes need to be optically thin which
implies that the mass of small dust grains (of order a micron or less) must be extremely low. 
Third, while most (pre\nobreakdashes-)transitional disks accrete onto
the central star at a rate which is lower than the median
accretion rate for TTS \citep[$\sim$10$^{-8}M_{\odot}\,yr^{-1}$;][\S~4.2]{hartmann98}, their rates are
still substantial (\S~4.2).  This indicates that considerable gas is located within the inner disk, although
we currently do not have many constraints on how this gas is spatially distributed.  
In the following sections, we review the clearing mechanisms that have been applied to
explain (pre\nobreakdashes-)transitional disk observations in light of the above constraints.

\bigskip
\noindent
\textbf{ 3.1 Non-dynamical Clearing Mechanisms}
\bigskip

A diverse set of physical mechanisms has been invoked to explain (pre\nobreakdashes-)transitional disk
observations, with varying levels of success.  Much 
of the attention is focused on dynamical interactions with one or 
more companion objects, although that subject will be addressed separately in 
\S~3.2.  There are a number of alternative scenarios that 
merit review here, including the effects of viscous evolution, particle growth 
and migration, and dispersal by (photoevaporative) winds.  

\bigskip
\noindent
\textit{ Viscous evolution }
\medskip

The interactions of gravitational and viscous torques comprise dominate
the structural evolution of disks over most of their lifetimes 
\citep{lynden74, hartmann98}.  An anomalous 
kinematic viscosity, presumably generated by MHD turbulence, drives a 
persistent inward flow of gas toward the central star \citep[e.g.,][]{hartmann06}.  
Angular momentum is transported outward in that process, resulting 
in some disk material being simultaneously spread out to large radii.  As time 
progresses, the disk mass and accretion rate steadily decline.  To first order, 
this evolution is self-similar, so there is no preferential scale for the 
depletion of disk material.  The nominal viscous evolution ($\sim$depletion) 
timescales at 10s of AU are long, comparable to the stellar host ages.

If the gas and dust were perfectly coupled, we would expect viscous evolution 
acting alone to first produce a slight enhancement in the FIR$/$mm SED 
(due to the spreading of material out to larger radii) and then settle into 
a slow, steady dimming across the SED (as densities decay).  Coupled with 
the sedimentation of disk solids, these effects are a reasonable explanation 
for the evolved disks (\S~2.1).  
That said, there is no reason to expect that viscous effects alone are capable 
of the preferential depletion of dust at small radii needed to produce the large
cleared regions typical of (pre\nobreakdashes-)transitional disks.
Even in the case of enhanced viscosity in the outer
wall due to the magneto-rotational instability \citep[MRI][]{chiang07}, this 
needs a pre\nobreakdashes-existing inner hole to have been formed via another mechanism in order to be effective.

\bigskip
\noindent
\textit{ Grain growth }
\medskip

The natural evolution of dust in a gas-rich environment offers two 
complementary avenues for producing the observable signatures of holes
and gaps in disks.  First is the actual removal of material due to the inward migration and 
subsequent accretion of dust particles.  This ``radial drift" occurs because 
thermal pressure causes the gas to orbit at subKeplerian rates, creating a 
drag force on the particles that saps their orbital energy and sends them 
spiraling in to the central star \citep{whipple72,weidenschilling77, brauer08}.  
Swept up in this particle flow, the reservoir of emitting grains 
in the inner disk can be sufficiently depleted to produce a telltale dip in the 
IR SED \citep[e.g.,][]{birnstiel12}.  A second process is related to the actual growth of dust 
grains.  Instead of a decrease in the dust densities, the inner disk only 
appears to be cleared due to a decrease in the grain emissivities: larger 
particles emit less efficiently \citep[e.g.,][]{dalessio01,draine06}.  Because 
growth timescales are short in the inner disk, the IR emission that 
traces those regions could be suppressed enough to again produce a dip in the 
SED.

The initial models of grain growth predicted a substantial reduction of small 
grains in the inner disk on short timescales, and therefore a disk clearing
signature in the IR SED \citep{dullemond05,tanaka05}.  More 
detailed models bear out those early predictions, even when processes that 
decrease the efficiency (e.g., fragmentation) are taken into account.  \citet{birnstiel12} 
demonstrated that such models can account for the 
IR SED deficit of transitional disks by tuning the local conditions so that 
small ($\sim$$\mu$m-sized) particles in the inner disk grow efficiently to 
mm$/$cm sizes.  However, those particles cannot grow much larger before their 
collisions become destructive: the resulting population of fragments would then 
produce sufficient emission to wash out the infrared SED dip.  The fundamental 
problem is that those large particles emit efficiently at mm$/$cm wavelengths, so 
these models do not simultaneously account for the ring-like emission 
morphologies observed with interferometers (Fig.~\ref{figmm}).  \citet{birnstiel12}  
argued that the dilemma of this conflicting relationship between 
growth$/$fragmentation and the IR$/$mm emission diagnostics means that particle 
evolution {\it alone} is not the underlying cause of cavities in disks.

A more in-depth discussion of these and other issues related to the 
observational signatures of grain growth and migration are addressed in the 
chapter by {\it Testi et al.}~in this volume.

\bigskip
\noindent
\textit{ Photoevaporation }
\medskip

Another mechanism for sculpting (pre\nobreakdashes-)transitional disk structures relies on 
the complementary interactions of viscous evolution, dust migration, and disk 
dispersal via photoevaporative winds 
\citep{hollenbach94,clarke01,alexander06,alexander07,gorti09a,gorti09b,owen10,owen11,rosotti13}.  
Here, the basic idea is that the high-energy irradiation of the disk surface by 
the central star can drive mass-loss in a wind that will eventually limit the 
re-supply of inner disk material from accretion flows.  Once that occurs, the 
inner disk can rapidly accrete onto the star, leaving behind a large (and 
potentially growing) hole at the disk center.  The detailed physics of this 
process can be quite complicated, and depend intimately on how the disk is 
irradiated.  The chapter in this volume by {\it Alexander et al.}~provides a 
more nuanced perspective on this process, as well as on the key observational 
features that support its presumably important role in disk evolution in 
general, and the (pre-)transitional disk phenomenon in particular.  

However, for the subsample of (pre\nobreakdashes-)transitional disks that have been studied in 
detail, the combination of large sizes and tenuous (but non-negligible) 
contents of the inner disks (\S~2.1, 2.2, 2.3), substantial accretion rates (\S~4.2), and relatively low 
X-ray luminosities (\S~4.3) indicate that photoevaporation does not seem to be a viable 
mechanism for the depletion of their inner disks 
\citep[e.g.,][]{alexander09,owen11,bae13b}.  In addition, there exist disks with very low accretion rates onto 
the star that do not show evidence of holes or gaps in the inner disk \citep{ingleby12}.

\bigskip
\noindent
\textbf{ 3.2 Dynamical Clearing by Companions}
\bigskip

Much of the theoretical work conducted to explain the clearings seen in disks
has focused on dynamical interactions with companions.
When a second gravitational source is present in the disk,
it can open a gap \citep[][see chapter by {\it Baruteau et al.} in this volume]{papaloizou07,crida06,kley12}.
The salient issue is whether this companion is a star or a planet.
It has been shown both theoretically and observationally that a stellar-mass companion can open a gap in a disk. 
For example, CoKu Tau$/$4 is surrounded by a transitional disk  \citep{dalessio05} and
it has a nearly equal mass companion with 
a separation of 8 AU \citep[\S~2.3;][]{ireland08}. Such a binary system 
is expected to open a cavity at 16--24~AU
\citep{artymowicz94}, which is consistent with the
observations \citep{dalessio05,nagel10}.  
Here we focus our efforts on discussing dynamical clearing by planets. 
A confirmed detection of a planet in a disk around a young star does not yet exist.  Therefore,
it is less clear if dynamical clearing by planets is at work in some
or most (pre\nobreakdashes-)transitional disks.
Given the challenges in detecting young planets in disks, the best we 
can do at present is theoretically explore what observational signatures
would be present if there were indeed planets in disks
and to test if this is consistent with what has been observed to date,
as we will do in the following subsections.

\bigskip
\noindent
\textit{Maintaining gas accretion across holes and gaps}
\medskip

A serious challenge for almost all theoretical disk clearing models posed to date
is the fact that some (pre\nobreakdashes-)transitional disks exhibit large dust clearings
while still maintaining significant gas accretion rates onto the star.
Compared to other disk clearing mechanisms (see \S~3.1), gap opening by planets can more
easily maintain gas accretion 
across the gap since
the gravitational force of a planet can ``pull''
the gas from the outer disk into the inner disk.

The gap's depth and the gas accretion rate across the gap are closely related. The disk accretion rate at any radius $R$ is defined as
$\dot{M}=2 \pi R\langle \Sigma v_{r}\rangle$, where $\Sigma$ and $v_{r}$ are the gas surface density and radial velocity at $R$. If we 
further assume  $\langle \Sigma v_{r}\rangle$=$\langle \Sigma\rangle$$\langle v_{r}\rangle$ and the accretion rate across the gap is a constant,
the flow velocity is accelerated by about a factor of 100 in a gap which is a factor of 100 in depth. 

In a slightly more realistic picture,  $\langle \Sigma v_{R}\rangle$=$\langle \Sigma\rangle$$\langle v_{R}\rangle$ breaks down within the gap
since the planet-disk interaction is highly asymmetric in the R-$\phi$ 2-D plane. Inside of the gap, the flow only 
moves radially at the turnover of the horseshoe orbit, which is very close to the planet. This high velocity flow can
interact with the circumplanetary material, shock with and accrete onto the circumplanetary disk, and eventually 
onto the protoplanet \citep{lubow99}. Due to the great complexity of this process, the ratio between the 
accretion onto the planet and the disk's accretion rate across the gap is unclear. Thus, we will parameterize 
this accretion efficiency onto the planet as 
 $\xi$. After passing the planet, the accretion rate onto the star is only $1-\xi$ of the 
accretion rate of the case where no planet is present.  Note that this parameterization assumes that
the planet mass is larger than the local disk mass so that the planet migration is slower than
the typical type II rate. For a disk with $\alpha=0.01$ and $\sim 10^{-8} M_{\odot}\,yr^{-1}$, the local
disk mass is 1.5 $M_{J}$ at 20 AU.  \citet{lubow06} carried out 3-D viscous hydrodynamic simulations   
 with a sink particle as the planet and found a $\xi$ of 0.9.
 \citet{zhu11} carried out 2-D viscous hydro-simulations, but depleted the circumplanetary
 material at different timescales, and found that $\xi$ can range between 0.1-0.9  depending 
 on the circumplanetary disk accretion timescale. 
The accretion efficiency onto the planet plays an essential role in the 
accretion rate onto the star. 

\bigskip
\noindent
\textit{Explaining IR SED deficits}
\medskip

Regardless of the accretion efficiency,
there is an intrinsic tension between significant gas accretion rates onto the star
and the optically thin inner disk region in (pre\nobreakdashes-)transitional disks. 
This is because the planet's influence on the accretion
flow of the disk is limited to the gap region \citep{crida06} whose outer and inner edge hardly differ by a factor of more than 2. After passing through the gap, the inner disk surface density is again
controlled only by the accretion rate and viscosity (e.g., MHD turbulence), similar to a full disk. Since a full disk's
inner disk produces strong NIR emission, it follows that transitional disks should also produce strong NIR emission, but they do not.

With this simple picture it is very difficult to explain  
transitional disks, which have strong NIR deficits, compared to pre\nobreakdashes-transitional disks.
For example, the transitional disk GM Aur has very weak NIR emission.  It has an optically thin inner disk at 10 $\mu$m with an optical depth of $\sim$0.01 and an
accretion rate of $\sim 10^{-8} M_{\odot}\,yr^{-1}$. Using a viscous disk model with $\alpha=0.01$, $\Sigma_{g}$ is derived to be 10--100
g/cm$^{2}$ at 0.1 AU. Considering that the nominal
opacity of ISM dust at 10 $\mu$m is 10 $cm^{2}/g$, the optical depth at 10 $\mu$m for the inner disk is 100--1000  \citep{zhu11},  which 
is 4--5 orders of magnitude larger than the optical depth ($\sim 0.01$) derived from observations.

In order to resolve the conflict between maintaining gas accretion across large holes and gaps while
explaining weak NIR emission, several approaches are possible.  In the following
sections, we will outline two of these, namely multiple giant planets
and dust filtration along with their observational signatures.

\bigskip
\noindent
\textit{A possible solution: multiple giant planets}
\medskip

\begin{figure}[htb]
\centering
 \includegraphics[width=8.5 cm]{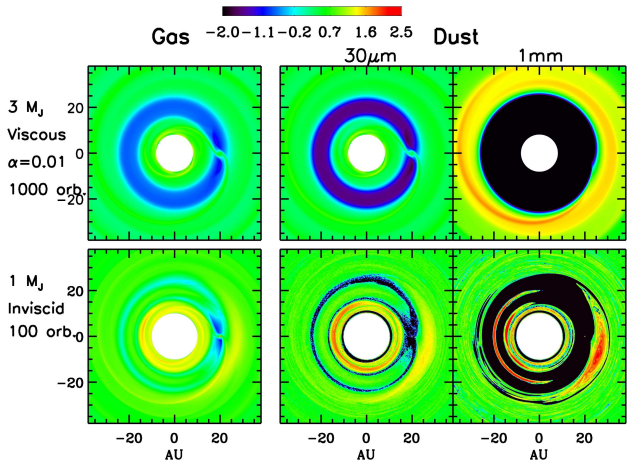}
 \vspace{-0.3 cm}
 \caption{\small
The disk surface density for the gas (left) and dust (right) in viscous ($\alpha$=0.01, top) and
inviscid (bottom) simulations. In the viscously accreting disk, the accretion flow carries small particles from
the outer disk to the inner disk while big particles are deposited at the gap edge. In the inviscid disk, particles are trapped in the vortex at the gap edge and the planet co-orbital region. Figure adapted from \citet[][]{zhu12} and {\em Zhu et al.}, in press.
\label{dynamic}}
 \end{figure}

One possibility is that multiple planets are present in (pre\nobreakdashes-)transitional disks.
If multiple planets from 0.1 AU
to tens of AU can open a mutual gap, the gas flow 
can be continuously accelerated and passes
from one planet to another so that a low disk surface density can sustain a substantial disk accretion rate onto the star  \citep{zhu11, dodson11}. 
However, hydrodynamical simulations have
shown that each planet pair in a multiple planet system will move into 2:1 mean motion resonance
\citep{pierens08}.  Even in a case with four giant planets, with the outermost one located at 20 AU, the mutual gap is from 2-20 AU \citep{zhu11}.  Therefore, to affect the gas flow at 0.1 AU, we need to invoke an even higher number of giant planets.

If there are multiple planets present in (pre\nobreakdashes-)transitional disks, the planet accretion 
efficiency parameter ($\xi$) 
cannot be large in order to maintain a moderate accretion rate onto the star. 
With $N$ planets in the disk, 
 the accretion rate onto the star
will be $(1-\xi)^{N}\dot{M}_{full}$. If $\xi=0.9$ and the full disk has a nominal accretion 
rate $\dot{M}_{full}=10^{-8} M_{\odot}\,yr^{-1}$, after passing two planets the accretion rate is 0.01$\times$
10$^{-8} M_{\odot}\,yr^{-1}$=10$^{-10} M_{\odot}\,yr^{-1}$ which is already below the observed accretion rates in (pre\nobreakdashes-)transitional disks (see \S~4.2).  On the other hand, if $\xi=0.1$, even with four
planets, the disk can still accrete at $(1-0.1)^{4}\dot{M}_{full}=0.66\dot{M}_{full}$. Thus,  
$\xi$ is one key parameter in the multi-planet scenario, which demands further study.

\bigskip
\noindent
\textit{Another possible solution: dust filtration}
\medskip

Another possibility is that the 
small dust grains in the inner disk are highly depleted by physical removal or grain growth,
to the point where the dust opacity in the NIR is far smaller than the ISM opacity.  
Generally,
this dust opacity depletion factor is $\sim$10$^{3}$ to 10$^{5}$  \citep{zhu11}. 
Dust filtration is a promising mechanism to deplete dust to such a degree.

Dust filtration relies on the fact that dust and gas in disks are not perfectly coupled.  Due to
gas drag, dust particles
will drift towards a pressure maximum in a disk \citep[see chapter by {\it Johansen et al.} in this volume;][]{weidenschilling77}.
The drift speed depends on the particle size, and particles with the dimensionless stopping time $T_{s}=t_{s}\Omega \sim 1$ drift fastest in disks.
If the particle is in the Epstein regime (when a particle is smaller than molecule mean-free-path) and has a density of
1 ${\rm g\, cm^{-3}}$, $T_{s}$ is related to the particle size, $s$, and the disk surface density, $\Sigma_{g}$, as
$T_{s}=1.55\times10^{-3}(s/1\, {\rm mm})(100\, {\rm g \,cm^{-2}}/\Sigma_{g})$ \citep{weidenschilling77}.
In the outer part of the disk (e.g., $\sim$50 AU) where $\Sigma_{g}\sim$ 1 or 10 g cm$^{-2}$, 1~cm particles
have $T_{s}\sim$ 1 or 0.1. Thus, submm and cm observations are ideal to reveal the effects of particle drift in disks.
At the outer edge of a planet-induced gap, where the pressure reaches a maximum, 
dust particles drift outwards, possibly overcoming their coupling to the inward accreting gas. 
Dust particles will then remain at the gap's outer edge while the gas flows through the gap. 
This process is called ``dust filtration'' \citep{rice06}, and
it depletes dust interior to the semi-major axis of the planet-induced gap, forming
a dust-depleted inner disk.

Dust trapping at the gap edge was first simulated by
\citet{paardekooper06,fouchet07}. However, without considering particle diffusion
due to disk turbulence, mm and cm sized particles will drift from tens of AU into the star within 200 orbits.
\citet{zhu12} included particle diffusion due to disk turbulence in 2-D simulations evolving over viscous timescales, where a quasi-steady state for both gas and dust has been achieved, and found that 
micron sized particles are difficult to filter by a Jupiter mass planet in a $\dot{M}=10^{-8} M_{\odot}\,yr^{-1}$
disk. \citet{pinilla12} has done 1-D calculations considering dust growth and dust fragmentation at the gap edge
and suggested that micron-sized
particles may also be filtered. Considering the flow pattern is highly asymmetric within the gap, 2-D simulations including
both dust growth and dust fragmentation may be
needed to better understand the dust filtration process. 

\begin{deluxetable}{llllll}
\tabletypesize{\small}
\tablecaption{Observable Characteristics of Proposed Disk Clearing Mechanisms}\label{tabchar}
\tablewidth{0pt}
\tablehead{
Mechanism			& Dust Distribution & Gas Distribution 	& Accretion Rate & Disk Mass	& L$_{X}$ }
\startdata
Viscous evolution  			& No hole$/$gap & No hole$/$gap & Low accretion & Low mass  & No dependence \\
Grain growth  			& No hole$/$gap & No hole$/$gap & Unchanged & All masses  & No dependence \\
Photoevaporation 		& $R_h$-radius hole & No$/$little gas within $R_h$	& No$/$low accretion 	& Low mass	& Correlated\\
$\sim$0.1~M$_J$ planet  	& Gap & No hole$/$gap 	& Unchanged & All masses  & No dependence \\
$\sim$1~M$_J$ planet 	& Gap & Gap 			& $\sim 0.1-0.9$ CTTS & Higher masses$^{a}$   & No dependence \\
Multiple giant planets 	     & Gap$/R_h$-radius hole & No$/$little gas within $R_h$ & No$/$low accretion & Higher masses$^{a}$ & No dependence
\enddata
\vskip -0.25in
\tablecomments{Here we refer to the dust and gas distribution of the inner disk.
Relative terms are in comparison to the properties of otherwise comparable 
disks around CTTS (e.g., objects of similar age, mass).
$^{a}$Higher mass disks may form planets easier according to core accretion theory (see chapter in this volume by
{\it Helled et al.}).  Observations are needed to test this.}
\end{deluxetable}

Dust filtration suggests that a deeper gap opened by a more massive planet 
can lead to the depletion of smaller dust particles. Thus
transitional disks may have a higher mass planet(s) than pre\nobreakdashes-transitional disks. 
Since a higher mass planet
exerts a stronger torque on the outer disk, it may slow down the accretion flow passing the planet 
and lead to a lower disk accretion rate onto the star. This is
consistent with observations that transitional disks have lower accretion rates than pre\nobreakdashes-transitional 
disks \citep{espaillat12,kim13}.
Furthermore, dust filtration can explain the 
 observed differences in the relative
distributions of micron sized dust grains
and mm sized dust grains.
Disks where the small dust grains appear
to closely trace the large dust distribution (i.e., category C disks from \S 2.3) could be in an advanced stage
of dust filtration, after
the dust grains in the inner disk have grown to larger sizes and can no longer
be detected in NIR scattered light. 
Disks with NIR scattered light imaging evidence for a significant amount of small, submicron sized
dust grains within the cavities in the large, mm-sized dust grains seen in the submm (i.e., the category A and C disks from \S 2.3), could be in an earlier stage of disk clearing via dust filtration.   

Future observations will further our understanding of the many physical processes which can occur when a giant planet is in a disk, such as dust growth in the
inner disk, dust growth at the gap edge \citep{pinilla12}, tidal stripping of the planets \citep{nayakshin13}, and planets in the dead zone \citep{morishima12,bae13}. 
In addition, here we have explored how to explain large holes and gaps in disks,
yet there may be smaller gaps present in disks (\S~2.1)
that may not be observable with current techniques.
The gap opening process is 
tied to the viscosity parameter assumed.
If the disk is inviscid, 
a very low mass perturber can also open a small
gap \citep[e.g., 10 M$_{\oplus}$;][]{li09,muto10,duffell12,zhu13,fung14}.  Disk ionization structure calculations have indeed
suggested the existence of such low turbulence regions 
in protoplanetary disks (i.e., the dead zone, see chapter by {\it Turner et al.} in this volume), 
which may indicate that gaps in disks are common.
Interestingly, in a low viscosity disk, vortices can also be produced at the gap edge \citep{li05}, 
which can efficiently trap dust particles \citep[Fig.~\ref{dynamic};][{\em Zhu et al.}, in press]{lyra09,ataiee13,lyra13} and
is observable with ALMA \citep{vandermarel13}. 
We note that other physical
processes can also induce vortices \citep[see chapter by {\it Turner et al.} in this volume;][]{klahr97,wolf02}, 
thus the presence of vortices cannot be
viewed as certain evidence for the presence of a planet within a disk gap.
In the near future, new observations with ALMA will provide
further constraints on the distributions of dust and gas in the disk, which will help
illuminate our understanding of the above points.

\bigskip
\section{\textbf{DEMOGRAPHICS}}\label{sec:demographics}
\bigskip

Basic unresolved questions underlie and motivate the study of disk demographics
around TTS.  
Do all disks go through a (pre\nobreakdashes-)transitional disk phase 
as stars evolve from disk-bearing stars to diskless stars?
Assuming an affirmative answer,  
we can measure a {\it transition timescale} 
by multiplying the fraction of disk sources 
with (pre\nobreakdashes-)transitional disk SEDs with the TTS disk lifetime \citep{skrutskie90}. 
Are (pre\nobreakdashes-)transitional disks created by a single process or multiple processes? 
Theory predicts that several mechanisms can generate 
(pre\nobreakdashes-)transitional disk SEDs, all of which are important for disk evolution.
Table~2 summarizes the generic properties expected for 
(pre\nobreakdashes-)transitional disks
produced by different mechanisms, 
as previously described in \S~\ref{sec:theory} and the literature 
\citep[e.g.,][]{najita07a,alexander07}.  
Which of these occur commonly (or efficiently) and on what timescale(s)? 
These questions can, in principle, be addressed demographically 
because different mechanisms are predicted to occur more readily in 
different kinds of systems (e.g., those of different disk masses and ages) and 
are expected to impact system parameters beyond the SED 
(e.g., the stellar accretion rate). We highlight developments along these lines below,
taking into account points raised in \S 2 and 3.

\begin{figure}[htb]
\begin{center}
 \includegraphics[width=8.2cm]{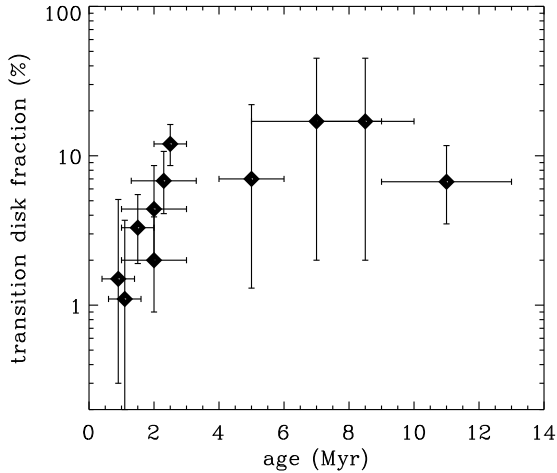}
 \end{center}
  \vspace{-0.3 cm}
 \caption{\small
The transitional disk fraction for ten star-forming regions and
young clusters as a function of mean stellar age.  Fractions were
estimated by taking the ratio of the number of objects
that exhibit a transitional disk SED with the total
number of disks in each region.  
Evolved disks and pre-transitional disks are not included here. 
The data are taken
from (in order of mean age): \citet{arnold12irs}, NGC 1333;
\citet{mcclure10}, Ophiuchus; \citet{furlan11}, Taurus;
\citet{manoj11}, Chamaeleon I; \citet{flaherty08},
NGC 2068/2071; \citet{oliveira10}, Serpens; \citet{muench07}, IC 348;
\citet{hernandez07a}, Orion OB1a/b; \citet{megeath05},
$\eta$ Chamaeleontis association; \citet{luhman12},
Upper Scorpius association.
\label{figfraction}}
 \end{figure}

\bigskip
\noindent
\textbf{ 4.1 Frequency and the Transition Timescale}
\bigskip

Because different clearing mechanisms likely operate on different timescales, 
demographic studies aim to identify the processes that produce 
(pre\nobreakdashes-)transitional disks in populations of different ages. 
One area that has been investigated thoroughly is the overall frequency of
transitional disks.
{\it Spitzer} 
surveys have had tremendous success in cataloging
young stellar object (YSO) populations, identifying dust emission from disks
around stars within 500~pc across the full stellar mass range and into
the brown dwarf regime at 3.6 to 8 $\mu$m, and for most of the stellar mass
range at 24 $\mu$m.  Based on the photometric SEDs, a number of studies 
have identified transitional disk candidates, allowing demographic comparisons
with stellar and disk properties. We note that reported transitional
disk frequencies to date should be taken as a lower limit on the frequency of holes and gaps
in disks given that, as noted in \S~2.1, it is harder to identify
pre-transitional disks based on photometry alone as well as smaller gaps in disks
with current data.

There has been some controversy regarding the
transition timescale, which as noted above can be estimated from the
frequency of transitional disks relative to the typical disk lifetime.
Early pre\nobreakdashes-{\it Spitzer} studies \citep[e.g.,][]{skrutskie90, wolk96}
suggested that the transition timescale was short, of order 10\%
of the total disk lifetime, a few $10^5$ years.  The values estimated from
{\it Spitzer} surveys of individual clusters$/$star forming regions
have a wider range, from as small as these initial estimates
\citep{cieza07,hernandez07b,luhman10,koepferl13}
to as long as a time comparable to the disk lifetime itself
\citep{currie09,sicilia09,currie11}.
Combining data from multiple regions, \citet{muzerolle10}
found evidence for an increase in the transitional disk frequency
as a function of mean cluster age.  

The discrepancies among these studies are largely the result of
differing definitions of what constitutes a disk in transition,
as well as how to estimate the total disk lifetime.
The shorter timescales are typically derived using more restrictive
selection criteria.  In particular, the evolved disks,
in which the IR excess is small at all observed wavelengths,
tend to be more common around older stars ($\gtrsim 3$ Myr) and low-mass
stars and brown dwarfs; the inclusion of this SED type will lead to
a larger transitional disk frequency and hence timescale for samples 
that are heavily weighted to these stellar types.
Moreover, the status of the evolved disks as {\it bona fide}
(pre\nobreakdashes-)transitional disks (i.e., a disk with inner clearing) is somewhat in dispute (see \S~2.1).
As pointed out by \citet{najita07a}, the different SED types
may indicate multiple evolutionary mechanisms for disks.
More complete measurements of disk masses, accretion rates,
and resolved observations of these objects
are needed to better define their properties and determine their
evolutionary status.

Combining data from multiple regions and restricting the selection
to transitional disk SEDs showing evidence for inner holes, \citet{muzerolle10}
found evidence for an increase in the transition frequency
as a function of mean cluster age.  However, there was considerable
uncertainty because of the small number statistics involved, especially
at older ages where the total disk frequency is typically 10\% or less.
Fig.~\ref{figfraction} shows the fraction of transitional disks relative to
the total disk population in a given region as a function
of the mean age of each region.  Here we supplement the statistics from
the IRAC$/$MIPS flux-limited photometric surveys listed in
\citet{muzerolle10} with new results from IRS studies of 
several star forming regions \citep{mcclure10,oliveira10,furlan11,manoj11,arnold12irs},
and a photometric survey of
the 11 Myr-old Upper Scorpius association \citep{luhman12}.
A weak age dependence remains, with frequencies of a few percent at
t~ $\lesssim$~2~Myr and $\sim$10 percent at older ages.  Some of this
correlation can be explained as a consequence of the cessation of
star formation in most regions after $\sim 3$ Myr;
as \citet{luhman10} noted, the frequency of transitional disks
relative to full disks in a region no longer producing new disks
should increase as the number of full disks decreases with time. 
Note that the sample selection does not include pre\nobreakdashes-transitional disks,
which cannot be reliably identified without MIR spectroscopy.
Interestingly, for the regions
with reasonably complete IRS observations (the four youngest
regions in Fig.~\ref{figfraction}), the transition fraction would increase
to $\sim 10-15$\% if the pre\nobreakdashes-transitional disks were included.
Note also that the IRS surveys represent the known membership of each region, and are reasonably complete down to spectral types of about M5.  We refer the reader to the cited articles for a full assessment of sample completeness.  All transitional disks in this combined sample are spectroscopically confirmed PMS stars.

\begin{figure}[htb]
\begin{center}
 \includegraphics[width=8.2cm]{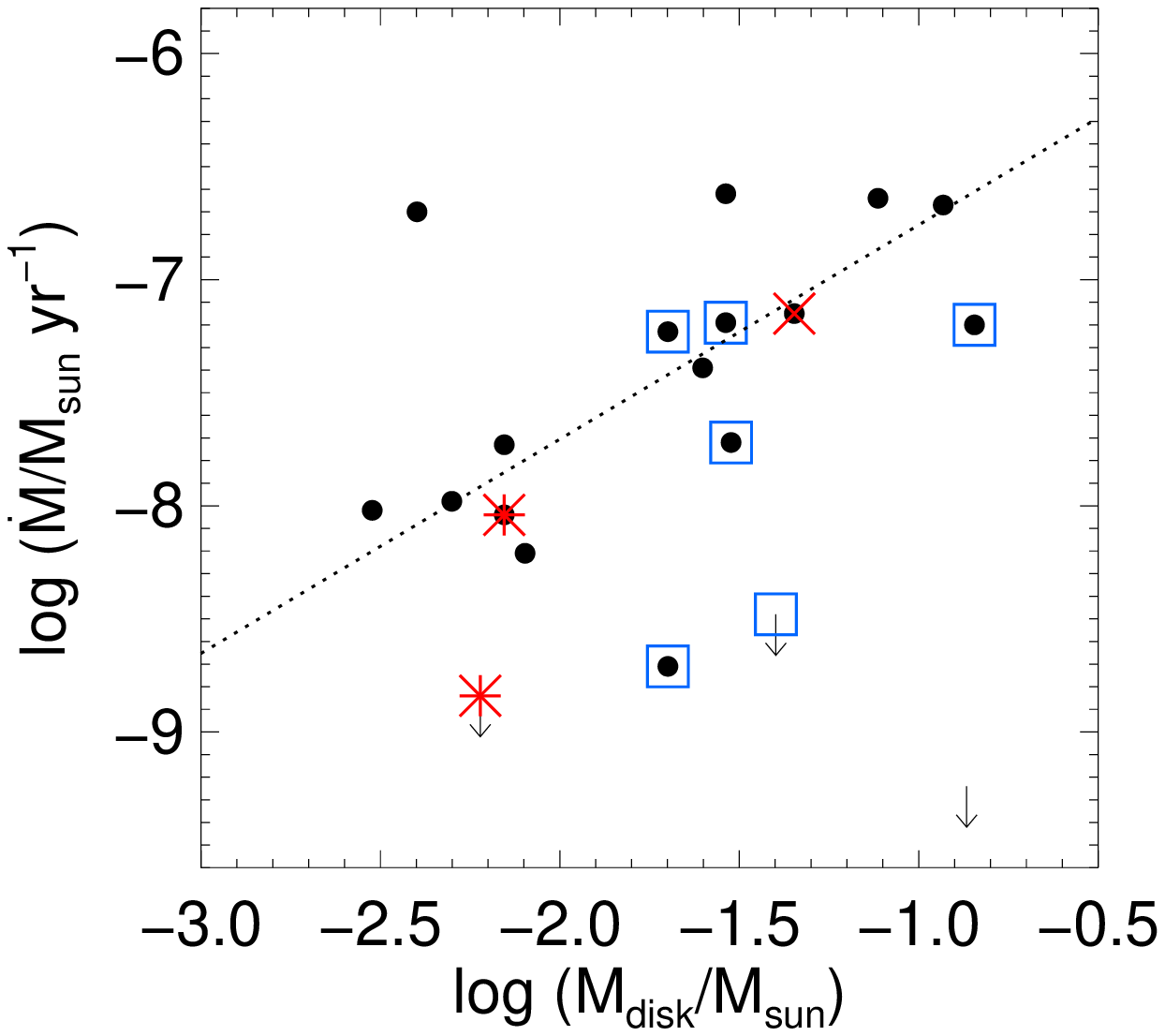}
  \end{center}
   \vspace{-0.3 cm}
 \caption{\small 
Preliminary \.M$_*$--M$_d$ for disks in Ophiuchus.
The dotted line indicates the locus occupied by CTTS in Taurus \citep{najita07a}.
Disks with submm cavities and (pre\nobreakdashes-)transitional disk SEDs are overlaid with a star (SR~21, bottom left; DoAr~44, middle);
the disk with a submm cavity but no obvious SED MIR deficit (SR~24S) is overlaid with an ``X'' .
Sources with A$_{V}$$>$14 \citep{mcclure10} are indicated with a blue box. 
Objects where only upper limits were available for the accretion rates are
indicated with an arrow (including SR~21, GSS~26, DoAr~25).
Accretion rates are from \citet{natta06} if available, otherwise 
from
\citet{eisner05}, and \citet{valenti93} (scaled). 
Disk masses are from \citet{andrews07,andrews10}.
}
\label{figmdotmass}
 \end{figure}

An important caveat to the above statistics is that SED morphology is
an imperfect tracer of disk structure and is most sensitive to large
structures that span many AU (see \S~2.1).  Smaller gaps
can be masked by emission from even tiny amounts of dust.
As mentioned in \S~2.2, resolved submillimeter imaging has
found a greater frequency of disks with inner dust cavities
among the 1-2 Myr-old stars in Taurus and Ophiuchus 
\citep[$\sim$30\%;][]{andrews11} than has been estimated from NIR flux deficits alone.
The statistics for older regions are incomplete, limited by weaker
(sub)millimeter emission that may be a result of
significant grain growth compared to younger disks
\citep{lee11,mathews12}.
Improved statistics await more sensitive observations such as
with ALMA.

\bigskip
\noindent
\textbf{ 4.2 Disk Masses and Accretion Rates}
\bigskip

Another powerful tool has been to compare 
stellar accretion rates and total disk mass.
Looking at full disks around CTTS and (pre\nobreakdashes-)transitional disks in Taurus, 
\citet{najita07a} found that compared to single CTTS, 
(pre\nobreakdashes-)transitional disks have stellar accretion rates $\sim 10$ times lower at the 
same disk mass and median disk masses $\sim 4$ times larger. 
These properties are roughly consistent with the expectations for Jupiter 
mass giant planet formation. 
Some of the low accretion rate, low disk mass sources could plausibly
be produced by UV photoevaporation, although none of the (pre\nobreakdashes-)transitional disks 
(GM Aur, DM Tau, LkCa 15, UX Tau A) are in this group. 
Notably, none of these disks were found to overlap the region of 
the \.M$_*$--M$_d$ plane occupied by most disks around CTTS. 

With the benefit of work in the literature, 
a preliminary \.M$_*$--M$_d$ plot can also be made for 
disks around CTTS and (pre\nobreakdashes-)transitional disks in Ophiuchus (Fig.~\ref{figmdotmass}). 
Most of the Oph sources are co-located with the Taurus CTTS, while 
several fall below this group. 
The latter includes DoAr~25 and the transitional disk SR~21 .
The other two sources have high extinction. Such sources ($A_v > 14$; Fig.~\ref{figmdotmass}, blue squares) 
are found to have lower \.M$_*$, on average, for their disk masses.
This is as might be expected, if scattered light leads to an underestimate of 
$A_v$, and therefore the line luminosity from which \.M$_*$ is derived. 
A source like DoAr~25, which has a very low accretion rate for its 
disk mass and also lacks an obvious hole based on the SED or submillimeter continuum 
imaging \citep{andrews08,andrews09,andrews10}, is interesting.
Does it have a smaller gap than can be measured with current techniques? 
Higher angular resolution observations of such sources could 
explore this possibility.

Like SR~21, both SR~24S and DoAr~44 have also been identified as having 
cavities in their submillimeter continuum \citep{andrews11}, 
although their accretion rates place them in the CTTS region of the plot. 
DoAr~44 has been previously identified as a pre\nobreakdashes-transitional disk based on its SED
\citep{espaillat10}. Whether SR~24S has a strong MIR deficit in its SED is unknown, 
as its SED has not been studied in as much detail. 

What accounts for the CTTS-like accretion rates of these systems? 
One possibility is that these systems may be undergoing low mass 
planet formation.
Several studies have described how low mass planets can 
potentially clear gaps in the dust disk while having little impact on 
the gas, with the effect more pronounced for the larger grains 
probed in the submillimeter \citep[see \S~3.2, also][]{paardekooper06,fouchet07}.

In related work, recent studies find that (pre\nobreakdashes-)transitional disks 
have lower stellar accretion rates on average than other disks 
in the same star forming region and that the accretion rates of pre\nobreakdashes-transitional disks are closer
to that of CTTS than the transitional disks \citep{espaillat12,kim13}.
However, other studies find little difference between the accretion 
rates of (pre\nobreakdashes-)transitional and other disks \citep[{\it Keane et al.,} submitted;][]{fang13}. 
This could be due to different sample selection
(i.e., colors vs. IRS spectra) and$/$or different methods for calculating accretion rates
(i.e., H$_{\alpha}$ vs. NIR emission lines).  
More work has to be done to better understand disk accretion
rates, ideally with a large IR and submm selected sample and consistent accretion
rate measurement methods.

\bigskip
\noindent
\textbf{ 4.3 Stellar Host Properties}
\bigskip

\begin{figure}[htb]
\begin{center}
 \includegraphics[width=8.2cm]{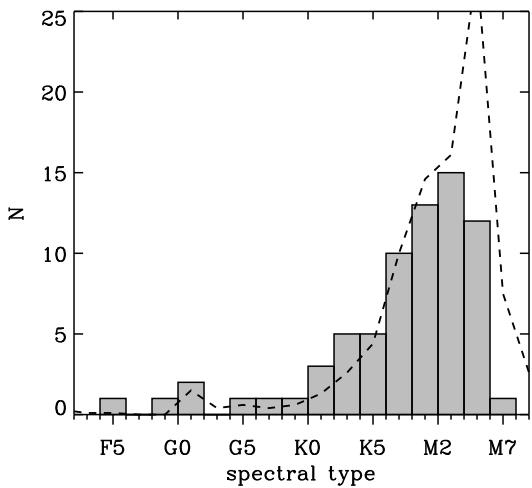}
 \end{center}
  \vspace{-0.3 cm}
 \caption{\small
The distribution of spectral types for the known 
transitional disks with 
spectral type F5 or later. The dashed line
indicates the combined parent sample of all disks in the 1--3~Myr old regions, 
scaled down by a factor of 10, highlighting the apparent deficit of 
transitional disks around stars later than $\sim$M4.
\label{figspt}}
 \end{figure}

One would naively expect an age progression from full disks to (pre\nobreakdashes-)transitional disks to diskless stars in a given region if these classes represent an evolutionary sequence.
Observational evidence for age differences, however, is decidedly mixed.
From improved parallax measurements in Taurus, \citet{bertout07} found
that the four stars surrounded by (pre\nobreakdashes-)transitional disks with age measurements
(DM Tau, GM Aur, LkCa~15, UX Tau~A) were intermediate between the mean CTTS
and weak TTS (i.e., stars that are not accreting) ages.  
However, the quoted age uncertainties are roughly equal to the ages of individual objects, and the small sample
is not statistically robust.  Moreover, studies of other regions have found
no statistical difference in ages between stars with and without disks
\citep[NGC~2264, Orion;][]{dahm05,jeffries11}.
In general, the uncertainties associated with determining stellar ages
remain too large, and the samples of (pre\nobreakdashes-)transitional disks too small, to allow any definitive
conclusion of systematic age differences for (pre\nobreakdashes-)transitional disks.

The frequency of transitional disks as a function of stellar mass
also provides some clues to their origins.
\citet{muzerolle10} showed that transitional disks
appeared to be underrepresented among mid- to late-M stars compared
with the typical initial mass function of young stellar clusters.
Using the expanded sample adopted for Fig.~\ref{figfraction}, this initial
finding appears to be robust (Fig.~\ref{figspt}).  The deficit remains even
after adding the known pre\nobreakdashes-transitional disks. 
However, the evolved disks (as opposed to the {\it transitional} disks) do appear to be much more common around lower-mass stars
\citep[e.g.,][]{sicilia08}.  As \citet{ercolano09} pointed
out, there may be a strong bias here as optically thick inner disks
around mid- to late-M stars produce less short-wavelength excess
and can be confused with true inner disk clearing.
Nevertheless, this discrepancy may provide another indication
of different clearing mechanisms operating
in different transitional disks.  For example, the relative dearth
of (pre\nobreakdashes-)transitional disks
around lower mass stars may reflect a decreased rate of giant planet
formation, as would be expected given their typically smaller
initial disk masses.

Among other stellar properties, the X-ray luminosity can provide further
useful constraints.  $L_X$ has been invoked as a major contributor
to theoretical photoevaporation rates \citep[e.g.,][]{gorti09a,owen12a},
as well as accretion rates related to MRI at the inner disk edge \citep{chiang07}.
Examining the (pre\nobreakdashes-)transitional disks in Taurus and Chamaleon I,
\citet{kim09} found correlations between
the size of the inner cleared regions and several stellar$/$disk properties
including stellar mass, X-ray luminosity, and mass accretion rate.
These possibly point to the MRI or photoevaporation mechanisms.
However, known independent correlations between stellar mass,
$L_X$, and mass accretion rate may also be responsible.
In a follow-up study with a large sample of 62 (pre\nobreakdashes-)transitional disks
in Orion, \citet{kim13} attempted to correct for the stellar
mass dependences and found no residual correlation between accretion
rate or $L_X$ and the size of the inner hole.
The measured properties of most (pre\nobreakdashes-)transitional disks
do not overlap with
the ranges of parameter space predicted by these models.
\citet{kim13} concluded that the demographics of
their sample are most consistent with giant planet formation
being the dominant process responsible for creating (pre\nobreakdashes-)transitional
disk SEDs.

To better understand the connection between underlying physical mechanisms and
observed disk demographics, more theoretical and observational work needs to be done.
Improved theoretical estimates on stellar ages, accretion efficiencies, and disk masses for different clearing mechanisms will aid in interpretation of the observations.
At the same time, larger ground-based surveys are essential to confidently constrain basic properties such as stellar accretion rates, ages, and disk masses.
In the near future, the higher resolution of ALMA will reveal the extent of gaps in disks, leading
to better statistics to measure the (pre\nobreakdashes-)transitional disk frequency and the transition timescale.
{\it JWST} will be key in addressing how the (pre\nobreakdashes-)transitional frequency depends on age.
The best constraints on disk demographics clearly require a multi-wavelength approach.

\bigskip
\section{\textbf{CONCLUDING REMARKS AND FUTURE\\ PROSPECTS}}
\bigskip

It is thought that practically all stars have planets (see chapter in this volume by
{\it Fischer et al.}).
If we start with the reasonable assumption that these planets formed out of disks,
then the question is not if there are planets in disks around
young stars, but what are their observable signatures.
To date we have detected large, optically thin holes and gaps in the
dust distribution of disks and
theoretical planet clearing mechanisms can account for some of their observed
properties.
These (pre\nobreakdashes-)transitional disks have captured the interest of many scientists since
they may be a key piece of evidence in answering one of the fundamental
questions in astronomy: how do planets form?

While other disk clearing mechanisms (e.g, grain growth, photoevaporation; \S~3.1)
should certainly be at play in most, if not all, disks \citep[and may even work more effectively
in concert with planets;][]{rosotti13}, here we speculate what kinds of planets could be clearing
(pre\nobreakdashes-)transitional disks, beginning with massive giant planets.
If a massive giant planet (or multiple planets) is forming in a disk, it will cause a large
cavity in the inner disk since it will be very efficient at cutting off the inner disk from the outer disk.
This will lead to significant depletion of small and large dust grains in the inner disk and lower accretion
onto the central star (\S~3.2).  This
is consistent with (pre-)transitional disks which have submm and NIR scattered light cavities and MIR deficits in the SED 
(e.g., LkCa~15, GM~Aur, Sz~91, RX J1604-2130; \S~2.1--2.3).
This is also consistent with the lower accretion rates measured for these objects (\S~4.2).
One caveat is that massive planets would be the easiest to detect, yet there has not been a
robust detection of a protoplanet in a (pre-)transitional disk yet.  

A less massive planet (or fewer planets) would still lead to substantial clearing
of the inner disk, but be less efficient at cutting off the inward flow of material from the outer disk (\S~3.2).
This is consistent with those (pre-)transitional disks that have a submm cavity but no MIR deficit in the SED, such as WSB~60, indicating that small dust still makes it through into the inner disk (\S~2.1--2.3).  
Interestingly, WSB~60 is in Oph and overall (pre\nobreakdashes-)transitional disks SEDs are rare in this region.
Given that Oph is quite young (age $\sim 1$\, Myr),
this may be indicative of the effects of dust evolution and$/$or planet growth. Disks in Oph could be in the initial stages of gap opening by planets
whereas in older systems, the dust grains in the inner disk have grown to larger sizes and$/$or the planets have grown large enough to more efficiently disrupt the flow of material into the inner disk.
The above suggests that most disks with planets go through a pre-transitional disk phase and that there are
many disks with small gaps that have escaped detection with currently available techniques.  More theoretical
work is needed to explore if all pre-transitional disks eventually enter a transitional disk phase as planets grow more massive and the inner disk is accreted onto the star.
Note also that there are some disks which have submm cavities and MIR SED deficits, but no evidence of clearing
in NIR scattered light (e.g., SR~21, DoAr~44, RX~J1615$-$3255).  Apparently, there is some important disk physics regarding accretion efficiencies that we are missing. It is necessary to study disks in other star forming regions with ALMA,
both to probe a range of ages and to increase the statistical sample size, 
in order to address questions of this kind.  Detections of protoplanets would be ideal
to test the link between planet mass and disk structure.

While much progress has been made understanding the nature
of the disks described above, future work may focus
more closely on disks with smaller optically thin gaps 
that could plausibly be created by low-mass planets.
The presence of low mass 
orbiting companions \citep[$\sim 0.1 M_J$;][]{paardekooper06} 
is expected to alter the dust disk significantly  
with little impact on the gas disk, 
whereas high mass companions ($\gtrsim 1 M_J$) can 
create gaps or holes in the gas disk and 
possibly alter its dynamics.
{\it Kepler} finds that super-earth mass objects are very common in 
mature planetary systems (see chapter in this volume by {\it Fischer et al.}), so it may be that smaller mass objects form fairly commonly in disks and open holes and gaps that are not cleared of dust. 
These systems are more difficult to identify with SED studies and current interferometers,
so potentially there are many more disks with gaps than currently known.
High spatial resolution images from ALMA will be able to detect these small gaps, in some cases even those with sizes down to about 3~AU \citep{wolf05,gonzalez12}. 

There are many remaining avenues that researchers can take to fully characterize clearing in disks around TTS.
Theoretically, we can simulate the influence of various mass
planets on different sized dust particles, which can be compared with observations at
various wavelengths to constrain a potential planet's mass.
We can use ALMA, VLTI, and JWST to test the full extent of disk holes and gaps as well as their frequency with respect to age.  
We can also make substantial progress studying the gas distributions in disks in the near future with ALMA,
particularly to
determine if the structure inferred for the disk from studies of the dust is the same as that in the gas.
Gas tracers may also reveal the presence of an orbiting planet via 
emission from a circumplanetary disk. 
Lastly, and perhaps most importantly, we need robust detections of protoplanets in disks around
young stars.
These future advances will help us understand how gas$/$ice giant planets and terrestrial planets form out of disks
and hopefully antiquate this review by the time of PPVII.\\


\textbf{ Acknowledgments.} The authors thank the anonymous referee, C. Dullemond, L. Hartmann, and L. Ingleby for insightful comments which helped improve this review.  Work by C.E. was performed in part under contract with Caltech funded by NASA through the Sagan Fellowship Program executed by the NASA Exoplanet Science Institute. Z.Z. acknowledges support by NASA through Hubble Fellowship grant
HST-HF-51333.01-A
awarded by the Space Telescope Science Institute, which is operated by the
Association
of Universities for Research in Astronomy, Inc., for NASA, under contract
NAS 5-26555.\\

Finally, a special recognition of the contribution of Paola D'Alessio, who passed away in November of 2013. She is greatly missed as a scientist, colleague, and friend.  

Among Paola's most important papers are those which addressed the issue of dust growth and mixing
in T Tauri disks.  In \citet{dalessio99}, she showed that disk models with
well-mixed dust with properties like that of the diffuse interstellar medium produced disks that
were too vertically thick and produced too much infrared excess, while lacking sufficient mm-wave
fluxes.  In the second paper of the series \citep{dalessio01} Paola 
and collaborators showed that
models with power-law size distributions of dust with maximum sizes around 1 mm produced much
better agreement with the mm and infrared emission of most T Tauri disks, but failed to exhibit
the 10 {\micron} silicate emission feature usually seen, leading to the conclusion that the large
grains must have settled closer to the midplane while leaving a population of small dust suspended
in the upper disk atmosphere.  This paper provided the first clear empirical evidence for 
the expected evolution of dust as a step in growing larger bodies in protoplanetary disks.
Another important result was the demonstration that once grain growth proceeds past sizes
comparable to the wavelengths of observation, the spectral index of the disk emission is
determined only by the size distribution of the dust, not its maximum.  Along with quantitative
calculations of dust properties, the results imply that the typical opacities used to estimate dust masses 
will generally lead to underestimates of the total solid mass present, a point that is frequently
forgotten or ignored.

In the third paper of this series \citep{dalessio06}, Paola and her collaborators
developed models which incorporated a thin central layer of large dust along with depleted upper
disk layers containing small dust.  The code which developed these models has been used in over 30
papers to compare with observations, especially those from the Infrared Spectrograph (IRS) 
\citep{furlan05,mcclure10}
 as well as the IRAC camera
\citep{allen04} on board the Spitzer Space Telescope, and more
recently from PACS on board
the Herschel Space Telescope \citep{mcclure12}, and from the
SMA \citep{qi11}.

Paola's models also played a crucial role in the recognition of transitional and pre-transitional
T Tauri disks \citep{calvet02,dalessio05,espaillat07b,espaillat10}.  
In many cases, particularly those of the pre-transitional
disks, finding evidence for an inner hole or gap from the spectral energy distribution depends
upon careful and detailed modeling.  The inference of gaps and holes from the SEDs is now
being increasingly confirmed directly by mm- and sub-mm imaging \citep{hughes09},
showing reasonable agreement in most cases with the hole sizes predicted by the models.
Combining imaging with SED modeling in the future will place additional constraints on the
properties of dust in protoplanetary disks.

Paola's influence in the community extended well beyond her direct contributions to the literature
in over 100 refereed papers.  She provided disk models for many other researchers as well as
detailed dust opacities.  The insight provided by her calculations informed many other
investigations, as studies of X-ray heating of protoplanetary disk atmospheres
\citep{glassgold04,glassgold07,glassgold09},
of the chemical structure of protoplanetary disks and propagation of 
high energy radiation 
\citep{fogel11,bethell11}
and on the photoevaporation of protoplanetary disks
\citep{ercolano08}. 

Paola's passing is a great loss to the star formation community and we remain grateful for her
contributions to our field.


\end{document}